\documentclass[11pt]{article}

\usepackage{amsmath,amssymb,amsfonts}
\usepackage{graphicx}
\usepackage{cite}
\usepackage{url}
\usepackage{hyperref}
\usepackage{color}
\usepackage{booktabs}
\usepackage{algorithm}
\usepackage{algorithmic}

\usepackage[margin=1in]{geometry}
\usepackage{setspace}
\title{A Predictive and Preventive Digital Twin Framework \\
for Indoor Wireless Networks}

\author{
    Jiunn-Tsair Chen \\
    WNC Corporation  \\
    jt.chen@wnc.com.tw
}

\date{} 

\begin{document}

\maketitle

\begin{abstract}
Wi-Fi networks increasingly suffer from performance degradation caused by contention-based channel access, dense deployments, and largely self-managed operation among mutually interfering access points (APs).  
In this paper, we propose a Digital Twin (DT) framework that captures the essential spatial and temporal characteristics of wireless channels and traffic patterns, enabling the prediction of likely future network scenarios while respecting physical constraints.

Leveraging this predictive capability, we introduce two analytically derived performance upper bounds—one based on Shannon capacity and the other on latency behavior under CSMA-CA (Carrier Sense Multiple Access with Collision Avoidance)—that can be evaluated efficiently without time-consuming network simulations.  
By applying importance sampling to DT-generated scenarios, potentially risky network conditions can be identified within large stochastic scenario spaces.

These same performance bounds are then used to proactively guide a gradient-based search for improved network configurations, with the objective of avoiding imminent performance degradation rather than pursuing globally optimal but fragile solutions.  
Simulation results demonstrate that the proposed approach can successfully predict time-dependent network congestion and mitigate it in advance, highlighting its potential for predictive and preventive Wi-Fi network management.
\end{abstract}


\section{Introduction}
Modern Wi-Fi networks have evolved into large-scale, dense, and highly heterogeneous systems.
In residential, enterprise, and public deployments, tens to hundreds of access points (APs) and stations (STAs) coexist within overlapping service areas, operating under limited spectrum resources and largely decentralized control.
While recent Wi-Fi standards have significantly improved peak data rates through wider bandwidth, higher-order modulation, and multi-antenna techniques, network robustness and latency performance remain fundamental challenges, especially under heavy and dynamic traffic conditions.

Unlike centrally scheduled cellular systems, Wi-Fi relies on CSMA/CA (Carrier Sense Multiple Access with Collision Avoidance), where medium access decisions are made independently by individual devices.
As a result, network performance is governed not only by physical-layer channel conditions, but also by complex interactions among traffic demand, QoS parameters, contention behavior, and spatial interference.
When these factors align unfavorably, Wi-Fi networks may experience abrupt performance degradation, manifested as excessive collisions, rapidly increasing frame delays, and unstable throughput.

Existing Wi-Fi optimization mechanisms are largely reactive.
They respond to performance degradation only after it has already occurred, typically through heuristic load balancing, admission control, or parameter tuning.
Such approaches are often insufficient in practice, because once a network enters a highly congested regime, recovery becomes difficult and user experience may already be severely impacted.
Moreover, many optimization decisions are made without a principled understanding of how close the network is to its fundamental performance limits.

To address these challenges, this paper proposes a predictive and preventive network optimization framework based on a Digital Twin (DT) of the Wi-Fi system.
Conceptually, the Digital Twin plays three distinct roles in this paper: (i) state abstraction, (ii) scenario generation, and (iii) risk screening. These roles are introduced progressively in the following sections.
The DT captures the essential spatial, temporal, and traffic characteristics of the physical network, and is used to generate a large ensemble of likely future network scenarios.
Rather than attempting to simulate every packet-level interaction, we focus on extracting analytical performance upper bounds that characterize the onset of network saturation and latency degradation.

Two complementary performance bounds are introduced.
The first is the Shannon capacity bound, which captures the fundamental limit imposed by wireless channel conditions.
The second is a latency-oriented bound expressed through normalized throughput, which characterizes the maximum sustainable offered load before contention and delay become unstable under CSMA/CA operation.
These bounds can be evaluated efficiently and do not require time-consuming network simulations.

\begin{figure}
    \centering
    \includegraphics[width=0.9\linewidth]{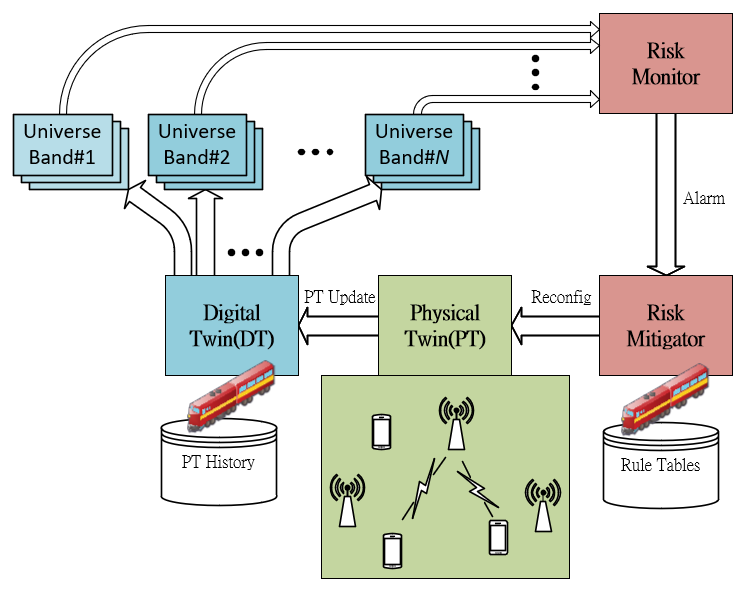}
    \caption{Block diagram of the proposed network optimization system}
    \label{fig:BlockDiagram}
\end{figure}

By combining the DT with importance sampling, the proposed framework can rapidly scan tens of thousands of predicted network conditions, identify scenarios that are likely to violate performance margins, and trigger early warnings.
When a risky scenario is detected, the same analytical bounds are reused to proactively search the configuration space—through AP association, load balancing, and QoS adaptation—to mitigate the predicted degradation before it occurs.  Putting these ideas together, we propose a predictive/preventive optimization system for Wi-Fi mesh 
network as shown in Figure \ref{fig:BlockDiagram}.  

The emphasis of this work is not on finding globally optimal configurations under idealized assumptions.
Instead, we aim to identify robust and feasible solutions that prevent catastrophic performance degradation, preserve low latency, and maintain stable operation under realistic uncertainty.
This philosophy aligns naturally with practical Wi-Fi deployments, where simplicity, predictability, and reliability often outweigh marginal gains in peak throughput.

Through analytical modeling and simulation-based validation, we demonstrate that the proposed framework can accurately predict impending network overload and effectively guide preventive reconfiguration.
Although the focus of this paper is on Wi-Fi systems, the underlying methodology is applicable to a broader class of distributed wireless networks where contention, latency, and uncertainty play a critical role.

\subsection{Previous Works and Their Short-Term Natures}

The challenges addressed in this paper lie at the intersection of wireless performance modeling~\cite{malone2011}, network optimization~\cite{viavi_ndt}, and emerging Digital Twin (DT) concepts~\cite{grieves2003, viavi_ndt}. Prior work in these areas has produced valuable tools and insights, but most existing approaches remain fundamentally short-term and reactive in nature. This subsection reviews representative prior efforts and clarifies the gap that motivates the predictive and preventive framework proposed here.

\paragraph{Analytical Modeling of Wi-Fi Performance.}
A substantial body of work has focused on analytical modeling of IEEE~802.11 MAC behavior, particularly under CSMA-CA operation. Classical Markov-chain–based models characterize throughput, collision probability, and delay under both saturated and non-saturated traffic conditions. Notably, Malone \emph{et al.}~\cite{Malone2007, malone2011} extended earlier saturated models to heterogeneous, non-saturated scenarios, providing accurate characterization of throughput and latency as functions of offered load and contention parameters. Such models are essential for understanding fundamental performance limits, but they are typically applied to instantaneous or static configurations rather than evolving network conditions.

\paragraph{Reactive Optimization and Heuristic Control.}
Many Wi-Fi optimization techniques—including load balancing, admission control, and QoS parameter tuning—are based on heuristic rules or local measurements. These approaches react to observed congestion after it occurs, for example by re-associating stations or adjusting contention parameters when throughput drops or delay rises. While effective in certain cases, they often lack a principled notion of feasibility: optimization is attempted even when the network is already operating beyond what is fundamentally achievable given its physical and contention constraints. Consequently, such methods may oscillate, converge slowly, or fail to restore acceptable performance.

\paragraph{Simulation-Driven and Data-Driven Approaches.}
System-level simulators such as ns-2 and ns-3~\cite{fall1999, ns3website} have been widely used to evaluate Wi-Fi behavior under complex scenarios. Although these tools offer high fidelity, they are computationally expensive and unsuitable for real-time or large-scale scenario scanning. More recently, data-driven and learning-based approaches have been explored, including reinforcement learning for channel selection, power control, and association decisions~\cite{arxiv_genai_dt}. These methods can adapt to observed conditions but typically require extensive training data, offer limited interpretability, and provide few guarantees when extrapolated beyond the training regime.

\paragraph{Digital Twins for Wireless Networks.}
The concept of Digital Twins has recently gained attention in wireless and networking research, particularly in the context of 5G and beyond~\cite{arxiv_genai_dt, viavi_ndt}. Many proposed wireless DTs focus on geometry-driven replication of the physical environment using ray tracing, site-specific channel models, or detailed radio maps derived from standards such as 3GPP~38.901~\cite{3GPP-38901}. These DTs are valuable for visualization, planning, and offline evaluation, but they primarily describe the present or hypothetical static state of the network. Temporal traffic evolution and contention-driven instability are often treated implicitly or ignored altogether.

\paragraph{Short-Term Nature of Existing Approaches.}
Across these strands of work, a common limitation emerges: most existing methods operate on short time horizons and react to degradation only after it has manifested. Analytical models are used to explain observed behavior rather than to anticipate future infeasibility; simulators evaluate specific scenarios rather than continuously scanning many possible futures; learning-based controllers adapt to feedback but do not explicitly reason about fundamental throughput–latency boundaries.

\paragraph{Positioning of This Work.}
In contrast, this paper emphasizes \emph{predictive and preventive} operation grounded in analytically derived performance upper bounds. Rather than optimizing instantaneous metrics, we explicitly characterize feasibility boundaries—through Shannon capacity and CSMA-CA latency bounds—and use them to assess how close a network is to collapse under predicted future conditions. By embedding these bounds within a Digital Twin that generates likely future scenarios, the proposed framework enables early detection of risky regimes and principled preventive intervention. This combination of analytical feasibility reasoning and forward-looking DT operation distinguishes the present work from prior reactive, short-term approaches.

\paragraph{Contributions}
The main contributions of this paper are summarized as follows:
\begin{itemize}
    \item We propose a \emph{predictive and preventive} Digital Twin (DT) framework for Wi-Fi networks, which anticipates likely future wireless and traffic scenarios based on historical observations from the physical twin, rather than reacting only after performance degradation has already occurred.
    \item We introduce two analytically derived and efficiently computable performance upper bounds—one based on Shannon capacity and the other based on latency behavior under CSMA-CA—which jointly serve as vendor-agnostic feasibility criteria for network performance monitoring and optimization.
    \item We develop an importance-sampling-based scenario generation mechanism within the DT, enabling the risk monitor to efficiently scan a large number of stochastic future scenarios and identify those that are most likely to cause imminent network performance degradation.
    \item We design a gradient-based risk mitigation procedure that reconfigures network parameters such as AP–STA associations and backhaul choices, aiming to restore sufficient performance margins rather than seeking globally optimal but fragile solutions.
    \item We validate the proposed framework through simulation of realistic household Wi-Fi mesh scenarios, demonstrating that the DT can successfully predict time-dependent performance degradation and proactively mitigate it before service quality deteriorates.
\end{itemize}

\paragraph{Organization of the paper}
Readers primarily interested in wireless channel and traffic modeling may focus on Section~\ref{sec:ChannelModel}, which introduces the spatial and temporal structures used in the Digital Twin.  
Sections~\ref{sec:ProposedSystems} and~\ref{sec:PerformanceUpperBounds} describe the proposed risk monitoring and mitigation framework together with the analytically derived performance upper bounds.  
Section~\ref{sec:Simulations} presents simulation scenarios and results that validate the predictive and preventive capabilities of the proposed approach.  
Readers interested mainly in practical implications and system-level insights may directly consult Sections~\ref{sec:ProposedSystems} and~\ref{sec:Simulations}.

\section{\label{sec:ChannelModel}Spatio-Temporal Digital Twin State and Observables}

This section defines the internal state representation of the Digital Twin, focusing on spatial signatures, temporal structure, and observable quantities inferred from normal network operation.  A Digital Twin capable of prediction and prevention must maintain an internal state that captures how the wireless network behaves across both space and time. This state cannot be a static snapshot, nor can it be a purely geometric reconstruction of the physical environment. Instead, it must encode those aspects of spatial structure and temporal evolution that directly influence network performance, while remaining compatible with the severe measurement constraints imposed by real deployments.

This section defines the spatio-temporal state representation adopted in this work. The emphasis is not on perfect realism, but on constructing a mathematically explicit and practically inferable representation that supports prediction, risk assessment, and preventive control.

\subsection{\label{Sec:SpaceSturcture}Space Structure} 

From the perspective of a wireless network, physical space is not directly observable. Walls, furniture, materials, and user locations do not appear explicitly in measurements. Instead, the environment manifests itself through received signal characteristics embedded in normal traffic. Among these, received power plays a central role.

Consider a service area covered by $N$ access points (APs). Without loss of generality, we focus on a single-input–single-output (SISO) setting for clarity; extensions to multiple antennas and beams follow the same principles. At any location, a station (STA) observes a vector of average received power levels
\begin{equation}
\mathbf{p} = [p_1, p_2, \ldots, p_N],
\end{equation}
where $p_n$ denotes the received power from AP $n$.

\begin{figure}
    \centering
    \includegraphics[width=1\linewidth]{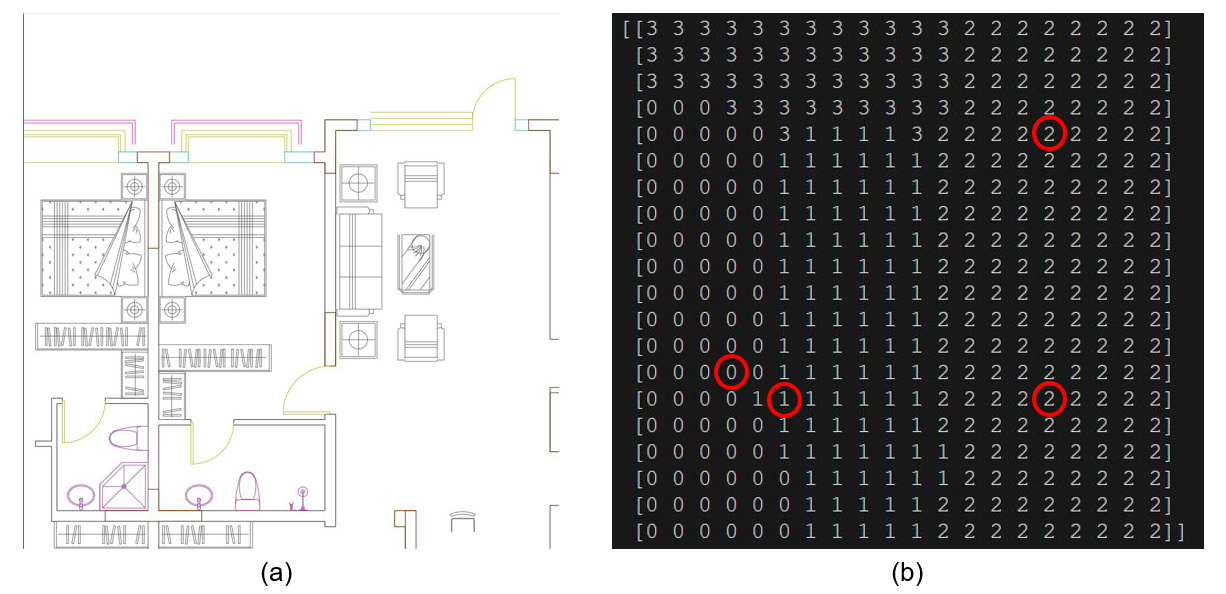}
    \caption{An example scenario to test spatial structure. (a)House layout, and (b)AP locations and its final spatial clustering. }
    \label{fig:HouseLayout}
\end{figure}

\begin{figure}
    \centering
    \includegraphics[width=1\linewidth]{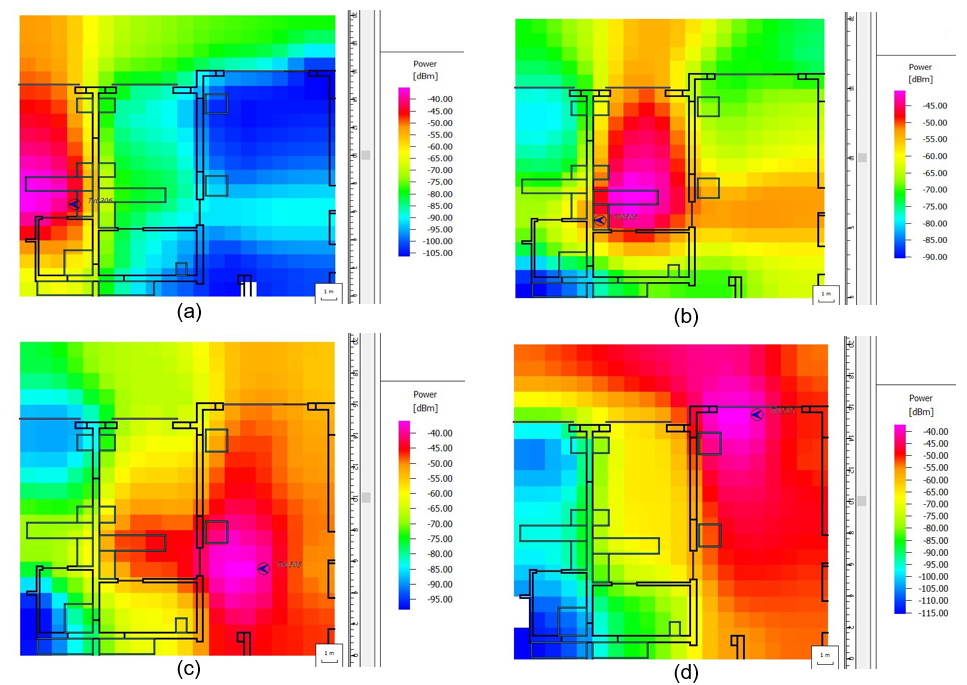}
    \caption{Heat maps of 4 APs at 2.4G frequency band.}
    \label{fig:HeatMaps}
\end{figure}

\begin{figure}
    \centering
    \includegraphics[width=1\linewidth]{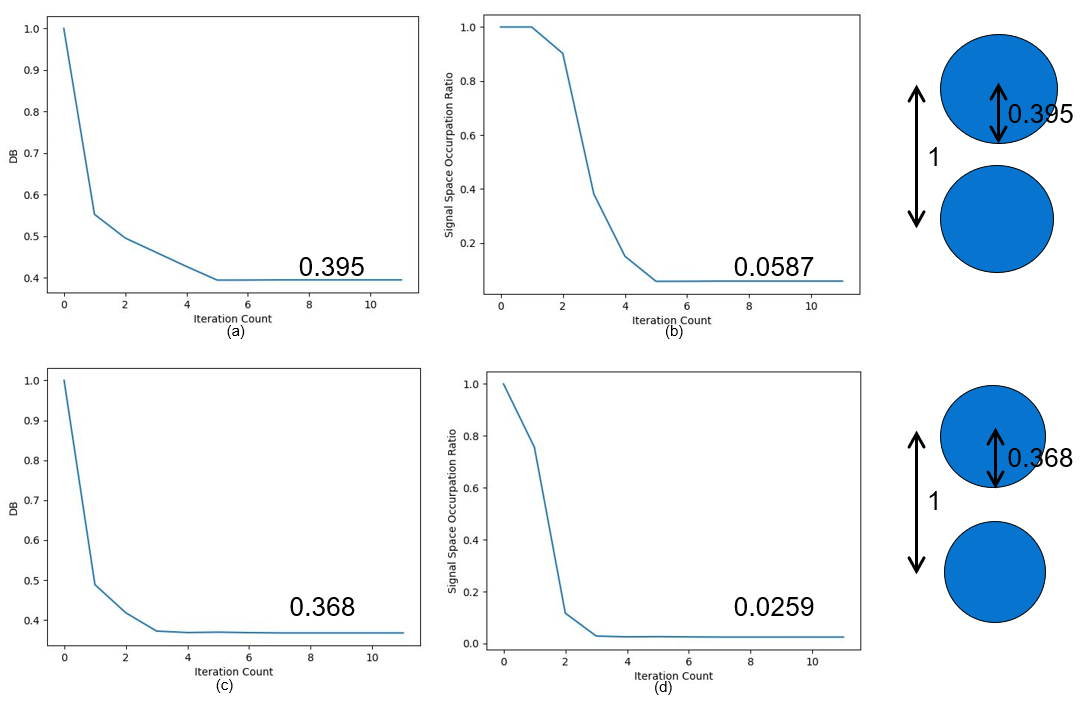}
    \caption{(a) and (b) are respectively DB index and cluster volume ratio of the example scenario with uniform user distribution, 
    (c) and (d) are respectively DB index and cluster volume ratio with non-uniform user distribution,  }
    \label{fig:DBMeasure}
\end{figure}

This vector is the network’s \emph{spatial signature} of that location. Crucially, $\mathbf{p}$ is not an arbitrary point in $\mathbb{R}^N$. Physical propagation laws, deployment geometry, and material interactions impose strong constraints on which power combinations can occur. As the dimensionality of the signature increases—for example through additional APs, frequency bands, or beam configurations—the ratio of physically feasible regions within the signal space shrinks dramatically. Valid signatures occupy a highly structured, low-dimensional subset of the ambient space.

This observation has two important consequences. First, spatial structure emerges naturally, even without explicit geometric modeling. Nearby locations tend to produce similar power signatures, while distant or obstructed locations differ in systematic ways. Second, clustering is not an algorithmic artifact, but a physical consequence of constrained signal propagation. Any Digital Twin that ignores this structure discards valuable information already present in the measurements.

For systems operating over multiple frequency bands or antenna configurations, the spatial signature generalizes to a higher-dimensional object. Let $B$ denote the number of frequency bands, $M_t^{(n)}$ the number of transmit antennas at AP $n$, and $M_r$ the number of receive antennas at the STA. The spatial signature can be expressed as
\begin{equation}
\mathcal{P} = \left[ \mathbf{P}_1, \mathbf{P}_2, \ldots, \mathbf{P}_N \right],
\end{equation}
where each $\mathbf{P}_n \in \mathbb{R}^{B \times M_t^{(n)} \times M_r}$ aggregates power responses across bands and antenna pairs. As dimensionality increases, the structure becomes richer and more discriminative, but also harder to infer from limited data.

\paragraph{Statistical Abstraction of Spatial Structure}

To make inference tractable, the Digital Twin represents spatial structure through statistical abstractions. A convenient first-order model is to view power signatures as drawn from a mixture distribution
\begin{equation}
f(\mathbf{p}) = \sum_{k=1}^{K} \pi_k \, \mathcal{N}(\mathbf{p}; \boldsymbol{\mu}_k, \boldsymbol{\Sigma}_k),
\end{equation}
where each component represents a region of similar propagation characteristics.

This abstraction is intentionally simplified. Real power signatures need not be Gaussian, and singularities or irregular boundaries may arise, particularly near transitions between regions. The purpose of this model is not to capture fine-grained physical detail, but to provide a structured statistical guide for a math-based teacher model to learn the dominant spatial patterns. More precise representations, such as radio-map–based formulations that explicitly model spatial fields, are possible and important, but are beyond the scope of this paper.

To demonstrate the spatial-structure property, we generate heatmaps for several APs in a 20m-by-20m two-dimensional
example household shown in Figure~\ref{fig:HouseLayout}(a).  The house partition and furniture locations are included.
After placing the APs, partition-material parameters (e.g., dielectric and conductive constants) are specified as
functions of operating frequency.  To mimic PT updates via simulation, we apply a ray-tracing algorithm to compute the
heatmap for each AP on a given operating band.  Figure~\ref{fig:HouseLayout}(b) shows the locations of four APs.  The
background numbers with a resolution of one square meter depict four clusters obtained by applying K-means clustering
to spatial power signatures.

Assuming three operating frequencies of 2.4G, 5G, and 6G~Hz, and a single-input-single-output (SISO) antenna system, we
obtain a 12-entry spatial power signature vector (4 APs $\times$ 3 bands) at each location in the household.  The heat
maps at 2.4G for the four APs are shown in Figure~\ref{fig:HeatMaps}.  Note that the clusters depicted in
Figure~\ref{fig:HouseLayout}(b) assume that STAs are uniformly distributed throughout the household.  Under this
assumption, clusters need not coincide with the service regions of OBSSs (Overlapping Basic Service Sets) associated
with each AP.  For example, AP\#2 (bottom right) and AP\#3 (top right) are in cluster 2, while there is no AP in
cluster 3.

Uniform STA distribution may be unrealistic and may suppress spatial structure, because some regions (e.g., inside
walls) do not permit user presence.  Alternatively, we may assume a non-uniform STA distribution where STAs are
uniformly distributed only over regions covering 1) all furniture and 2) within a one-meter vicinity of furniture,
since furniture locations are where people tend to appear.  In Figure~\ref{fig:DBMeasure}, we compute the DB (Davies--Bouldin) index and
the cluster volume ratio (in the 12-dimensional space) over K-means iterations, under both uniform and non-uniform STA distributions.   The DB index is defined as
    \begin{equation}
        {\rm DB} = \frac{1}{n} \sum_{i=1}^n \max_{j \neq i} \frac{\sigma_i + \sigma_j}{d(c_i, c_j)},
    \end{equation}
    where $n$ is the number of clusters, $c_i$ is the centroid of cluster $i$, $\sigma_i$ is the average distance of
    all elements in cluster $i$ to centroid $c_i$, and $d(c_i, c_j)$ is the distance between centroids $c_i$ and $c_j$.
Since clusters separate well, only a few iterations are required for convergence.  In both cases,
clustering is clear; moreover, the DB index under non-uniform STA distribution is even lower than that under uniform
distribution, as expected.

This example can be readily extended to multiple-input-multiple-output (MIMO) array systems, which greatly increase
the dimension of spatial power signatures.  In such cases, clustering effects are expected to become more dominant,
especially at higher frequency bands where signals decay faster and angular resolution of propagation paths is higher.

\subsection{\label{Sec:TimeStructure}Time Structure}

Spatial structure alone is insufficient for prediction. Wireless networks are dynamic systems whose behavior evolves over time in response to user activity, traffic demand, and protocol interactions. Treating time merely as a simulation index or averaging window obscures the very correlations needed for foresight. 
Rather than modeling traffic as a single stochastic time series, the Digital Twin represents user activity through multiple latent on/off processes operating at different time scales.
For each user and access category (AC), activity is determined by several independent causes, such as transaction execution, application-level tasks, and user presence or availability.

Each cause is modeled as an alternating on/off process defined by two duration parameters: the time the process remains active once it turns on, and the time it remains inactive once it turns off. These durations are not assumed to be fixed or memoryless. Instead, they evolve over time and reflect long-term changes in user behavior.

To capture this evolution without resorting to packet-level simulation or rigid parametric assumptions, the Digital Twin predicts the duration parameters themselves. For each user, AC, and activity cause, the on- and off-durations are treated as latent random processes whose mean values are tracked online using lightweight state estimators. The resulting duration sequences are then algorithmically expanded into minute-level on/off activity traces.

Temporal regularities such as daily and weekly routines are incorporated through piecewise-stationary parameter regimes. The day is partitioned into fixed multi-hour intervals and the week into weekday and weekend periods, with each regime maintaining its own duration estimates. This structure allows the Digital Twin to represent long-term behavioral patterns while remaining compatible with limited and noisy observations.

The final activity state for a given user and AC is obtained as the logical conjunction of all latent processes. This reflects the intuitive constraint that traffic can be generated only when all necessary conditions—such as user presence and application intent—are simultaneously satisfied.
Conditioned on activity, packet arrivals are modeled using simple memoryless processes parameterized by access category.

\paragraph{Observables and Measurement Constraints}

The Digital Twin operates under strict observability constraints. It does not assume access to explicit user locations, full channel state information, or external sensing infrastructure. Instead, it relies on measurements naturally available during normal operation: received power embedded in traffic, aggregate traffic statistics, and coarse timing information.

Both spatial and temporal structures are therefore latent. They must be inferred indirectly from incomplete, noisy, and unevenly sampled data. Any attempt at prediction without acknowledging these constraints is fundamentally unrealistic.

\paragraph{Assumptions, Bias, and the Role of Teacher Models}

The spatio-temporal models described above impose assumptions that introduce bias. This bias is not a flaw of the framework, but an unavoidable consequence of reality. The diversity of realistic indoor environments—including layouts, materials, AP placements, frequency bands, user locations, traffic patterns, and QoS demands—defines a space of possibilities that is far too large to be exhaustively measured or calibrated.

In this context, math-based teacher models play a crucial role. They provide structured supervision that captures first-order spatial and temporal regularities, enabling learning-based student models to be trained in the absence of comprehensive data. These student models are then expected to refine and correct bias through interaction with the real environment.

\paragraph{Implications for the Digital Twin Framework}

By defining the Digital Twin state as a structured spatio-temporal process inferred from traffic-driven observations, this framework shifts the focus from replication to reasoning. The goal is not to reproduce physical reality in detail, but to maintain a state that is sufficient for predicting performance evolution and assessing risk.

This state representation provides the foundation upon which the throughput–latency feasibility analysis and predictive Digital Twin operation are built. With the state defined, Section \ref{sec:PerformanceUpperBounds} can then turn to the mathematical core that determines what performance is fundamentally achievable.

\begin{figure}
    \centering
    \includegraphics[width=1\linewidth]{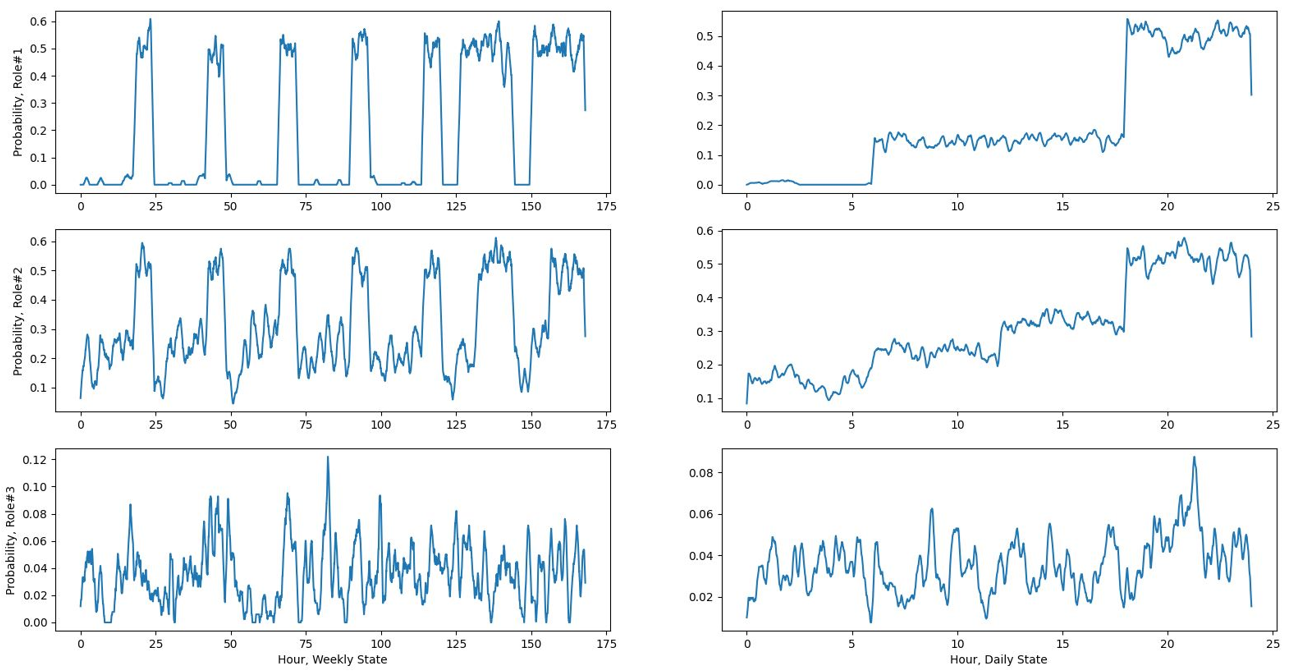}
    \caption{The probability of turning active for 3 roles (left column) and their respective long-term daily average (right column). }
    \label{fig:TrafficProbability}
\end{figure}

\begin{figure}
    \centering
    \includegraphics[width=1\linewidth]{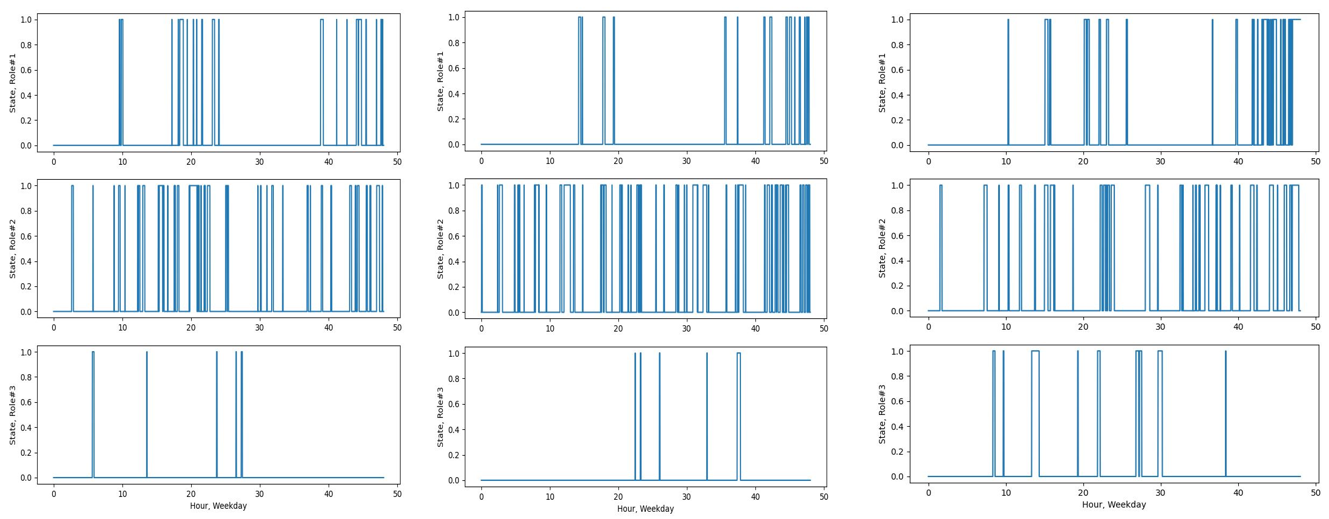}
    \caption{Three random traffic realizations shown as 3 columns for 3 roles.}
    \label{fig:TrafficRealization}
\end{figure}

\begin{figure}
    \centering
    \includegraphics[width=1\linewidth]{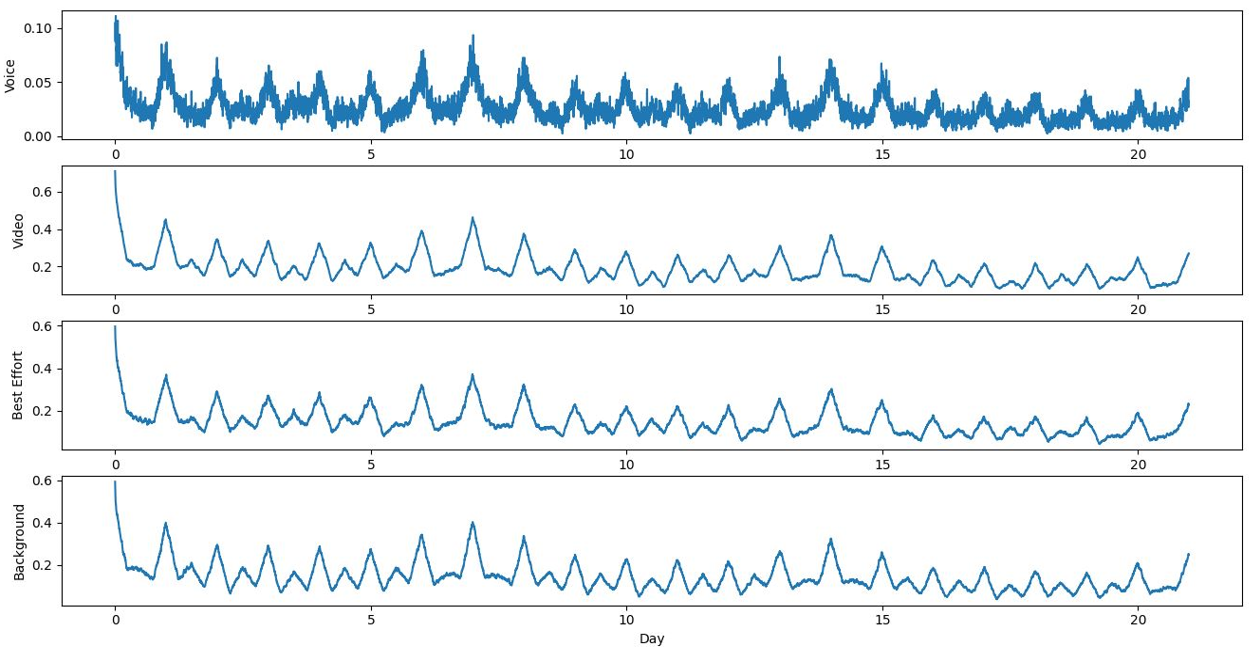}
    \caption{Temporal auto-correlation of 4 AC (Access Category) traffic.}
    \label{fig:TimeCorrelation}
\end{figure}

To illustrate the effects of time structure, we construct an example household scenario with three roles:
an office worker who stays at home at night and during weekends, a college student who may sometimes stay home during
the daytime, and a random visitor or an auto-running device that may become active at any time once in a while.
Without enumerating all implementation details, each role is assigned behavior patterns with different combinations of AC (Access Category) traffic,
and Poisson processes are used to emulate their generated traffic.

The left-column subfigures of Figure~\ref{fig:TrafficProbability} show the activation probabilities of each role over a
week of 168 hours.
The bottom subfigure appears random and has a much lower presence probability.
The right-column subfigures show the corresponding 24-hour daily averages.
The top two daily averages reveal that daytime traffic corresponds to the mean of 1) less busy weekday traffic and
2) busier weekend traffic.
Therefore, daily statistics alone are insufficient to capture role-based traffic behavior, since some patterns repeat
on a weekly basis.

The three columns of Figure~\ref{fig:TrafficRealization} show three independent traffic realizations from the three
roles.
The top two rows illustrate coherent role behavior over a two-weekday period.

Finally, computing the auto-correlation of the traffic demands for each AC and summing demands over all roles yields
Figure~\ref{fig:TimeCorrelation}.
Traffic is separated by AC because different ACs can have very different mean arrival times and mean service durations.
Regardless of AC type, strong daily correlation is observed, and correlation fades gradually as day separation increases.
In addition, a longer periodic correlation of one week is also evident.
These figures motivate the use of structured, multi-timescale temporal abstractions for future traffic prediction in the Digital Twin.

\section{\label{sec:ProposedSystems}Proposed Predictive and Preventive DT Framework}

Because time-dependent performance degradation and its root causes are often difficult to identify,
one effective way to address such problems is to leverage information contained in historical data.
Consider a scenario in which people in a household tend to gather in the living room to chat on weekend nights,
which repeatedly overloads the AP serving that area.  After the residents report the issue to the operator,
a technician may be dispatched to the site during a weekday daytime, which is typically the most convenient time.
However, because the problem does not manifest under those conditions, no abnormality is observed in the running Wi-Fi system,
and the same performance degradation continues to occur on subsequent weekend nights.

\begin{figure}
    \centering
    \includegraphics[width=1\linewidth]{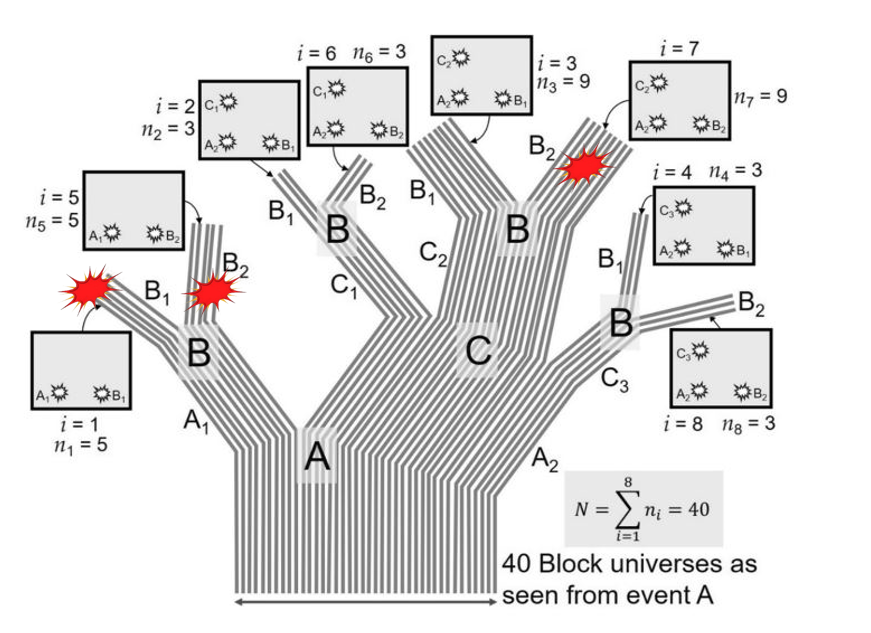}
    \caption{The concept of parallel universes. \cite{ParallelUniverses}}
    \label{fig:ParallelUniverse}
\end{figure}

The proposed solution is intended to address exactly this type of problem.  Based on historical data, the proposed
system is expected to trigger alarms whenever similar events that lead to performance degradation are predicted to occur.
Upon receiving an alarm, the risk mitigator automatically searches for and applies a better network configuration,
thereby enabling predictive and preventive network management. 

The proposed framework consists of three functional components: 
(i) future scenario generation in the Digital Twin, 
(ii) a risk monitor that scans predicted scenarios, and 
(iii) a risk mitigator that proactively searches for feasible reconfiguration.

\subsection{Future Prediction in Digital Twins}
We next examine Figure~\ref{fig:BlockDiagram} in more detail in this and the next two subsections.
The DT is responsible for generating network scenarios that may lead to performance degradation, while ensuring these scenarios
do not violate physical laws and remain likely to occur, so that the risk monitor can periodically scan them.
For example, the DT may predict one hour into the future once every three hours.  In that case, the DT first updates its network
scenario using a snapshot from the PT, and then evolves the scenario forward for one hour.

Future prediction in the Digital Twin focuses on the evolution of activity parameters rather than direct prediction of traffic states. At each prediction cycle, the Digital Twin updates the latent duration parameters associated with each user, access category, and activity cause.

Using the most recent parameter estimates, the Digital Twin extrapolates on- and off-duration sequences over the prediction horizon. These sequences are expanded into minute-level activity indicators and combined across activity causes using logical conjunction. Traffic arrivals are then generated conditionally on activity using simple stochastic models appropriate for each access category.

This approach avoids explicit modeling of packet-level dynamics while preserving the dominant temporal structure relevant for performance feasibility analysis. Uncertainty enters through the evolution of the duration parameters, which are continuously updated as new observations become available. As a result, future scenarios are statistically plausible but not deterministic, enabling effective risk screening without excessive computational cost.

Multiple future scenarios are generated to account for uncertainty in parameter evolution. Scanning a larger number of realizations reduces the likelihood of missing rare but critical congestion events, while the performance bounds introduced later allow these scenarios to be evaluated efficiently.

\subsection{\label{sec:RiskMonitor}Risk Monitor}

The risk monitor triggers an alarm to notify the risk mitigator that an impending risk exists, i.e., network performance may
soon degrade to a level that is unsatisfactory for some users, either due to insufficient throughput or excessive transmission
latency.  It is therefore necessary to identify potential problems and to define rigorously what constitutes unsatisfactory service.
Different ACs have different service-quality expectations.  Wi-Fi defines QoS parameters such as the minimum contention window $W_0$,
backoff depth $m$, waiting time defined by AIFSN, and transmission duration defined by TXOP (IEEE 802.11e), intended to meet
these service requirements.  

  To begin, in this subsection we first consider natural performance bounds that are fundamental and should not be overlooked.
We then summarize service requirements for the four AC levels, define conditions that justify raising alarms, and use those rules
to guide the design of the risk monitor.

In a wireless network, several natural performance bounds exist and can be summarized as follows:
\begin{enumerate}
    \item Throughput: the Shannon capacity gives the maximum achievable throughput over a channel as a function of received power,
    noise-plus-interference level, and allocated bandwidth, with a margin reserved to account for implementation penalties.  The
    required margin depends on cost, time, and effort devoted to implementation.  For example, low-end systems may not justify
    heavy optimization, and therefore require larger margins when estimating achievable throughput.
    
    \item Latency: under CSMA-CA, there exists an implicit upper limit on the number of users sharing the channel before latency becomes
    unacceptable for a given QoS demand.  This limit can be characterized through collision probability and latency in multi-rate,
    multi-QoS scenarios (saturated or not).  These results will be presented in Section~\ref{sec:LatencyBounds}.  Note that the throughput
    in the previous item assumes no contention.  As will be shown in Section~\ref{sec:LatencyBounds}, when CSMA-CA is adopted,
    both latency and throughput degrade as the number of users increases beyond a manageable range.
    
    \item Jitter: based on transmission error rate, jitter can be anticipated.  Frequent retransmissions and missed decoding deadlines
    increase jitter probability.  Jitter is more likely when
        A) diversity is low (e.g., a 2T2R system transmitting 2 spatial streams),
        B) STAs are mobile (causing channel fluctuation), and
        C) the environment varies quickly (e.g., a party or a busy shopping mall).
    These causes depend on capabilities and operating modes of the involved APs and STAs.
\end{enumerate}
All of these measures can be used by the risk monitor as alarm indicators.  However, performance degradation due to insufficient
throughput and excessive latency is typically more significant and more directly observable than jitter.  Therefore, in this paper
we focus on the first two measures, whose formulations will be given in Section~\ref{sec:PerformanceUpperBounds}.

We aim to provide sufficient resources to each AP and STA so that expected service quality can be maintained.
Accordingly, we do not model detailed device-driver operations and do not attempt instantaneous reactions to fast network variations.
Instead, we rely on analytically derived performance measures, introduce an implementation margin, and optimize these measures.
In other words, we do not explicitly model traffic shaping, frame scheduling, and related mechanisms, which typically require
complicated simulations/experiments and involve too many parameters simultaneously.  We assume network deployment and device drivers
are sufficiently competent to avoid resource waste that would cause performance far below the feasible bounds derived here.
With this approach, the proposed algorithm is implementation-neutral and can operate with Wi-Fi products from any vendor.
For high-end products we can reduce the implementation margin; for low-end products we expand it.

For the four AC levels, the corresponding service requirements are summarized as follows:
\begin{enumerate}
    \item Voice ({\rm VO}): For VoIP calls, latency of 150--300~ms is acceptable, and the data rate is typically 100~kbps.
    \item Video ({\rm VI}): For video streaming, live streaming may allow up to 10~s of delay, while low-latency applications may
    require latency below 100~ms.  Typical data rates depend on resolution: HD video requires about 5~Mbps, while 4K video requires about 25~Mbps.
    \item Best Effort ({\rm BE}) and Background ({\rm BK}): These are used for email, web browsing, and large file transfers.
    Latency is less meaningful here (e.g., downloading a 100~Gbit file at 100~Mbps takes over 15 minutes for completion), but users
    still expect sufficiently high data rates to perceive stable progress.  We therefore use 25~Mbps as a satisfaction criterion for these ACs.
\end{enumerate}
To simplify these criteria, in the remainder of this paper we refer to frame latency below 100~ms as negligible.
The risk monitor should raise an alarm when
\begin{enumerate}
    \item VO traffic cannot maintain 100~kbps with negligible delay,
    \item VI traffic cannot maintain 5~Mbps with negligible delay, and
    \item either BE or BK traffic cannot maintain 25~Mbps.
\end{enumerate}
Note that dumping large volumes of BE/BK traffic onto the channel is generally undesirable: it provides little benefit to
those transfers while increasing collisions that harm higher-priority traffic.  We therefore assume Wi-Fi devices are smart
enough to apply traffic shaping or scheduling for BE/BK under such conditions.  In that case, although the end-to-end completion
time for bulk transfers may remain long, individual frames can still experience negligible latency without harming higher-priority traffic.
In this paper, our goal is to ensure that allocated resources are sufficient and that at least one feasible mechanism exists
to maintain QoS.  This keeps the proposed solution independent of specific AP/STA models and vendors.  It is then the vendors'
responsibility to implement MAC/PHY mechanisms that achieve these feasible targets, analogous to selecting rates or MCS based on
a Shannon-bound estimate with an implementation margin, where the coding/modulation implementation is responsible for delivering
the feasible performance.  Because of implementation penalties and idealizations in derivations, alarm criteria may be relaxed by
introducing additional margin.

The risk monitor must scan scenarios efficiently to cover a large space and reduce the probability of detection misses.
In contrast, once an alarm is raised, the risk mitigator must search for a reconfiguration that eliminates the impending degradation.
Such optimization is not expected to occur frequently, but when it does, it may require substantially more computation to explore
and validate candidate solutions.  This is the intended division of responsibility: the risk monitor must be fast with wide coverage,
while the risk mitigator must be more accurate and analyze the problem more deeply.

\subsection{\label{sec:RiskMitigator}Risk Mitigator}

In a Wi-Fi service area, sufficient wireless resources are typically available to serve APs and STAs; since it serves no value in forcing them to compete aggressively for scarce resources.  Under such conditions, even a suboptimal resource allocation
may be adequate.  Searching for a globally optimal allocation with significant effort, only to apply it briefly before network
conditions change, may not be worthwhile.  Nevertheless, some scenarios may cause the network to become stuck in a poor allocation
if decisions are made using fixed heuristic rules.  For example, if each STA always associates with the AP that yields the highest
RCPI, then when many STAs gather near one AP, that AP becomes overloaded.  Offloading some STAs to neighboring APs can balance load
and mitigate the problem.  In the risk mitigator, we seek feasible and improved (but not necessarily optimal) solutions whenever
an alarm is raised.  Detecting unsatisfactory service and quickly locating a practical mitigation is more important than pursuing
a global optimum that yields only marginal gains not perceived by users.

With this objective, we perform gradient search in the design space using cost functions derived in
Section~\ref{sec:PerformanceUpperBounds}.  A large number of adjustable parameters are available in Wi-Fi specifications that can
redistribute resources, including:
\begin{enumerate}
    \item re-associating a STA from one AP to another AP, 
    \item selecting a different channel number (within the same or a different band) to reduce mutual interference among APs,
    \item re-configuring backhaul to improve efficiency,
    \item optimizing QoS parameters in each OBSS: minimum contention window $W_0$, contention layer $m$, wait period AIFSN, and TXOP,
    \item adaptively controlling transmit power at APs and STAs,
    \item optimizing CCA level to allow higher interference and reduce collision probability, 
    \item optimizing OBSS-PD (packet detection) level to enable simultaneous transmissions across OBSSs (e.g., under cell coloring),
\end{enumerate}
as well as many other applicable parameters in the Wi-Fi specification.

\begin{figure}
    \centering
    \includegraphics[width=0.6\linewidth]{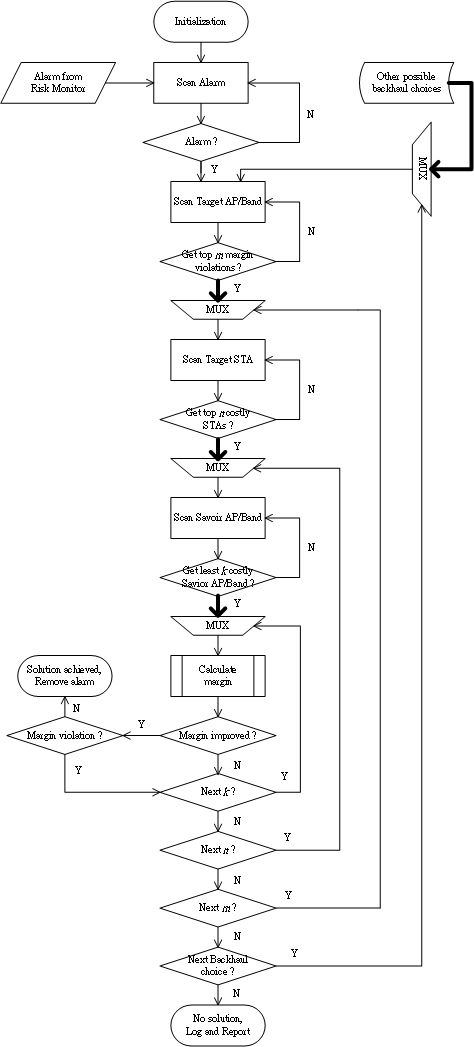}
    \caption{Flow chart of searching for solution search in the risk mitigator.}
    \label{fig:GradientSearch}
\end{figure}

As a proof of concept, in this paper we focus on the first three items above.
We perform gradient search by iterating through these three dimensions using three nested search loops, as shown in
Figure~\ref{fig:GradientSearch}.  The main idea is as follows:
\begin{enumerate}
    \item If resource deficiency occurs in an OBSS, the AP in that OBSS should divert the STAs that consume the most wireless
    resources to other APs.  We refer to this AP and those STAs as the target AP and target STAs.
    \item We then identify another AP that can take over service to the target STAs before they are dropped by the target AP.
    We refer to this AP as a savior AP.
    \item If a savior AP is found and begins to serve the target STAs, it may itself become resource-deficient.
    In that case, the savior AP becomes the next target AP.  We therefore iterate across APs to ensure that no AP remains
    resource-deficient before claiming that a feasible solution has been found.
    \item If no solution is found, we may escalate to search for a better backhaul configuration.  Backhaul optimization is treated
    as a last resort for the following reasons:
    \begin{enumerate}
        \item It can induce large-scale configuration changes that may severely disrupt ongoing service,
        \item frequent switching among backhaul configurations may be inefficient, since configurations can differ significantly,
        \item the search space can be large, and many candidates are obviously unreasonable and not worth exploring.
    \end{enumerate}
    \item After completing the procedure above, it is still possible that no feasible solution exists.  In that case, the scenario
    should be logged and reported, and additional measures may be required, such as deploying additional APs or physically relocating APs.
\end{enumerate}

\begin{figure}
    \centering
    \includegraphics[width=1.0\linewidth]{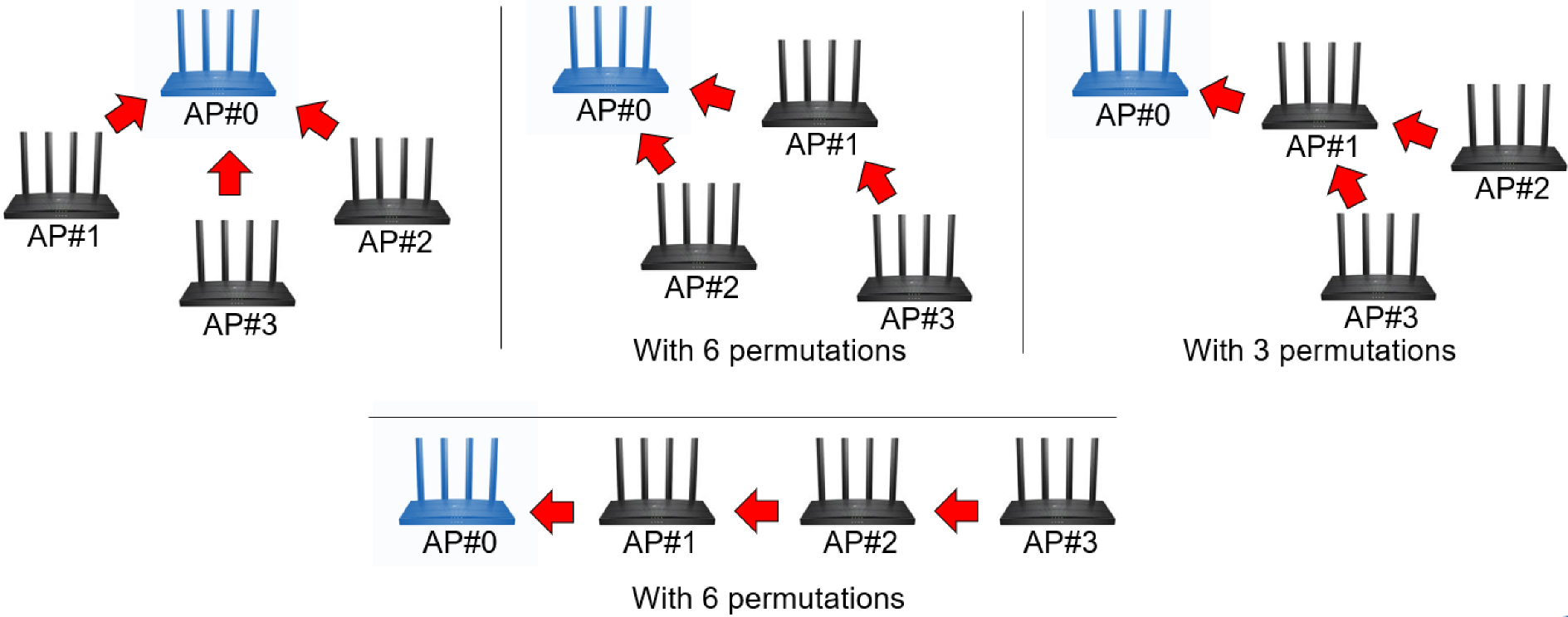}
    \caption{All possible backhaul connections in a service area with 4 APs}
    \label{fig:BackhaulSearch}
\end{figure}

We next elaborate on solution search over different backhaul configurations.
Assume four APs are connected wirelessly with a designated controller AP\#0, and there are $K$ STAs.
All possible backhaul connections (excluding MLO (multi-link operation)) are depicted in Figure~\ref{fig:BackhaulSearch}.
Since APs connect wirelessly, there are up to $16 \times 3^3$ possible backhaul configurations to search, assuming three frequency bands.
It is also too costly to perform exhaustive search over the $12^K$ possibilities for associating $K$ STAs to the 12 combinations of (AP, band).
However, most permutations are not reasonable in practice; for example, an AP typically would not choose a far-away AP to build backhaul
when a nearer AP is available.  Consequently, only a small subset of configurations are sensible.  We therefore pre-select a short list of
reasonable backhaul choices and store them in a table for searching, as shown in Figure~\ref{fig:GradientSearch}.

When searching for a savior AP for a STA, the STA must select both an AP and a frequency band.
For AP selection, one may choose the AP with the best 2.4~GHz received power, because (ignoring interference) receive power tends
to decrease with increasing carrier frequency, and relative RCPI ordering at 2.4~GHz often correlates with other bands.
For band selection, one may choose the highest frequency band that still provides sufficient received power, since higher bands offer
more channel numbers and may reduce interference.  However, each AP typically selects only one channel within each band and one bandwidth
choice (20/40/80/160~MHz).  Thus, choosing a higher band does not automatically increase available resources, but may reduce interference.
Conversely, if interference is not a concern (e.g., in a stand-alone house), the 2.4~GHz band may be preferred due to better coverage.

In calculating frame latency, we do not directly manipulate raw throughput on each link and the detailed traffic load from each STA/AC.
Throughput and offered load affect latency through received power and the selected MCS.  It is therefore convenient to
1) define normalized throughput as the fraction of time the channel is occupied by payload transmission, abstracting away PHY rate details; and
2) define normalized offered load as the fraction of time there is payload ready for transmission from each STA/AC.
Throughput and offered load relate to their normalized counterparts via a transmission-efficiency factor (bits per second), which depends on
MCS, coding rate, MAC/PHY header overhead, and inter-frame gaps and backoff behavior under CSMA-CA.
When evaluating overall normalized throughput, we do not distinguish uplink from downlink, because all traffic of the same AC
(from the AP or from STAs) contends for the same wireless resources.  As will be shown in the next section, before severe collisions
occur as traffic increases, neither uplink nor downlink experiences significant degradation.

The gradient solution search shown in Figure~\ref{fig:GradientSearch} can be summarized as follows:
\begin{enumerate}
    \item{\bf AP Alarm:} Among the AP/bands whose normalized throughput margins are below a threshold, identify the AP/band with the $m$ lowest margin as AP/band-Target.
    \item{\bf STA Scan:} Among the STAs served by AP/band-Target, identify the costliest $n$ STAs that are assigned the worst data rate as STA-Targets.
    \item{\bf AP Scan:} Among candidate AP/bands that can serve each STA-Target, identify $k$ AP/bands that consume the least normalized throughput to serve STA-Target,
    while keeping normalized throughput low enough to improve the overall minimum margin across AP/bands, and select AP/band-Savior.
    \item{\bf AP-STA Matching:} Switch STA-Target from AP/band-Target to AP/band-Savior.
    \item{\bf Matching Loop:} Loop through 1)--4) until either no AP/band violates the normalized throughput margin or no further improvement is possible.
    \item{\bf Backhaul Loop:} If any AP/band still violates the margin, select a different backhaul configuration from a pre-defined short list and loop through 1)--5)
    until either no AP/band violates the margin or no further improvement is possible.
    \item{\bf Report:} If any AP/band still violates the margin, log the scenario and seek additional resources.
\end{enumerate} 

In conclusion, we summarize the key ideas underlying the DT, risk monitor, and risk mitigator.
It is always possible to define infinitely many scenarios that can break a given Wi-Fi network.  The key question is whether such scenarios
obey physical laws and occur with non-negligible probability, since it is not useful to solve problems that are unlikely to occur.

The role of the Digital Twin is not to simulate packet-level behavior, but to emit statistically plausible future scenarios. To act on these scenarios, the system requires fast, implementation-agnostic criteria to determine whether a scenario is feasible or risky. This motivates the performance upper bounds introduced next.

\section{\label{sec:PerformanceUpperBounds}Performance Upper Bounds} 
\subsubsection*{Overview and Design Philosophy}
The performance bounds introduced in this section are intended as \emph{feasibility screens} rather than exact predictors of packet-level behavior.
They are designed to identify operating regimes in which no reasonable configuration can satisfy throughput or latency requirements, independent of specific MAC or PHY implementations.
This choice prioritizes analytical tractability, computational efficiency, and robustness under uncertainty, enabling fast scanning of large numbers of predicted scenarios in the Digital Twin.

\subsubsection*{Why Upper Bounds Instead of Simulation}
Packet-level simulation is computationally expensive and ill-suited for scanning the large number of stochastic future scenarios generated by the Digital Twin.  
Analytical upper bounds provide fast, implementation-agnostic feasibility screens that are sufficient for early risk detection and preventive reconfiguration, which is the primary objective of this framework.

\subsubsection*{Two Suggested Bounds}
A learning network requires cost functions, either (1) within a neural network for gradient-based solution search or
(2) within an optimization framework for system-level optimization,
we adopt two network performance upper bounds for this purpose, since they can be computed accurately and efficiently:
\begin{enumerate}
    \item {\bf Data throughput:} the Shannon bound under power resources indicated by RCPI and/or beamforming, and 
    \item {\bf Transmit latency:} the time between when a frame becomes ready for transmission and when it is
    successfully transmitted after going through the exponential backoff mechanism.
\end{enumerate}
It is our intention to consider only performance upper bounds and ignore implementation details, so that the proposed
system can be broadly applicable to APs and STAs from different vendors.  

\subsection{Throughput Bounds}

We reserve a minimum average throughput for each STA/AC depending on its AC level, as roughly defined in
Section~\ref{sec:RiskMonitor}.  A few points should be noted:
\begin{enumerate}
    \item The throughput requirement is defined in terms of its \emph{average} value.  That is, if the instantaneous
    throughput demand of a STA/AC fluctuates significantly around this average, we do not treat it as a problem here.
    The involved AP/STA is responsible for applying traffic smoothing or other mechanisms to handle such fluctuations.
    \item Multiple STAs may share the same wireless channel under various multiple-access schemes.  In this subsection, we allow
    any multiple-access scheme as long as the Shannon bound is not exceeded.  However, in the next subsection on transmit latency,
    we specifically assume the multiple-access scheme to be CSMA-CA, since it is the primary cause of frame latency in Wi-Fi systems.
    Therefore, the throughput bound derived in this subsection is also applicable to non-Wi-Fi systems, such as 5G mobile systems.
\end{enumerate}

If the operational bandwidth in an OBSS is announced and fixed, each STA accessing this channel bandwidth can be equivalently
treated as consuming some fraction of that bandwidth resource, i.e., sharing a ``pie'' among STAs regardless of the specific
multiple-access mechanism.  As long as this pie is not fully consumed, we can define a percentage resource margin for each AP.
We aim to maximize the minimum margin across all APs in the service area.  We implement this objective using the following steps:
\begin{enumerate}
    \item Define the minimum acceptable throughput of each associated STA/AC as its offered load based on its QoS level.
    Sum both uplink and downlink offered loads over all STA/ACs, since they contend for reciprocal channels in the same
    channel resource pool.
    \item Add protocol overhead due to headers and waiting gaps.  If the average frame size is unavailable, the maximum
    transmission unit (MTU) is assumed to estimate header and gap overhead, since we are computing an upper bound.
    \item For each target AP and each of its frequency bands, use the received power and the rank of the traffic channels
    (or those inferred from the beacon RCPI) at each STA/AC associated with the target AP/band to compute the minimum bandwidth
    required to satisfy each STA/AC's offered load under the Shannon bound.  Subtract the required bandwidth from the total
    available bandwidth (e.g., 20~MHz) of the target AP/band.
    \item If the leftover bandwidth of any AP/band is below a pre-defined margin (e.g., 20\%), then an alarm is triggered and the
    risk mitigator begins gradient or exhaustive search for a better network configuration.  The same upper-bound calculation
    above is used as the cost function to increase the minimum throughput margin.
\end{enumerate}
Note that STAs farther from their associated AP generally consume more bandwidth than those nearby.  Therefore, an AP will
naturally prefer serving STAs with stronger received power from that AP, and conversely, a STA will prefer associating with APs
that provide stronger received power.

\subsection{\label{sec:LatencyBounds}Latency Bounds}

In an OBSS running CSMA-CA, we first compute the frame delay of each STA/AC in the OBSS.
The Markov chain shown in Figure~\ref{fig:MarkovChart} represents such a scenario.
By solving the steady-state probability of each state in the Markov chain, described by a system of nonlinear equations,
we can derive the normalized throughput, the frame collision probability, the transmission attempt probability,
and the average latency of data frames for each STA/AC.  If there are $K$ STA/ACs, we solve a system of $3K$ nonlinear
equations with unknowns $q_l$, $p_l$, and $\tau_l$, which respectively represent the data availability rate
(or equivalently the normalized offered load), the collision probability, and the transmission attempt probability
of the $l^{\rm th}$ STA/AC.

\subsubsection{Markov Chain}
\begin{figure}
    \centering
    \includegraphics[width=1.0\linewidth]{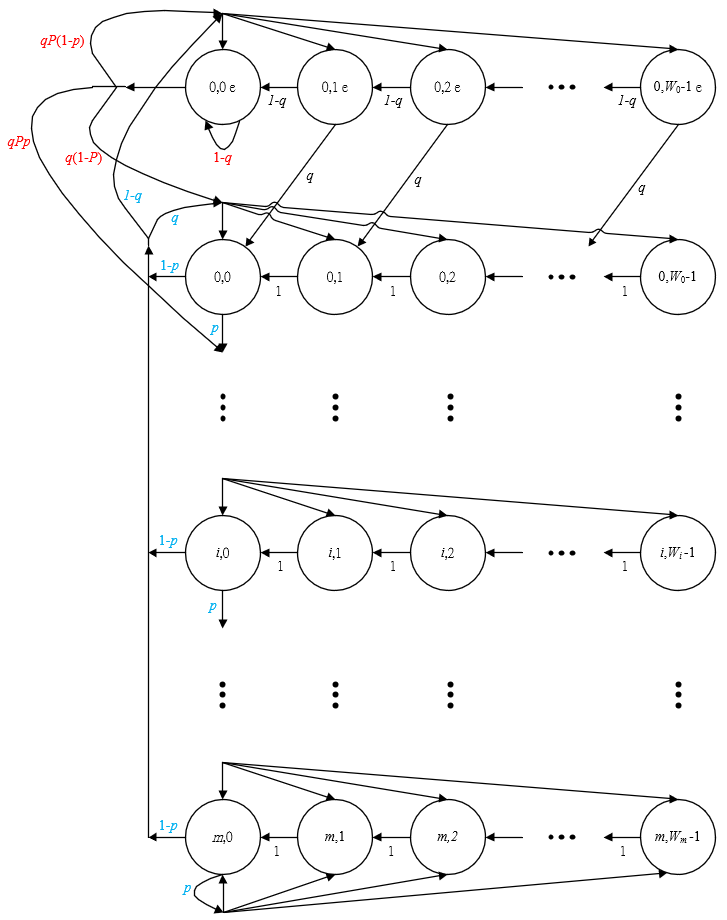}
    \caption{A Markov chain of CSMA-CA with non-saturated traffic}
    \label{fig:MarkovChart}
\end{figure}

Let us first describe the Markov chain shown in Figure~\ref{fig:MarkovChart}. 
Part of materials in this sub-section are summarized from~\cite{Malone2007} for the clear description of this paper .
The top row denotes the states when there is no data frame to transmit.
The remaining $m$ rows, from the second row to the last row, denote the states in the backoff stages.
Whenever a frame collision happens, the state moves from the current row to a state in the next row below, until it reaches the bottom row.
After each collision, the backoff level $i$ increases by one and saturates at its maximum value $m$.
That is, the states in the $(i+2)^{\rm th}$ row are the countdown states after $i$ collisions, for $0 \le i < m$.

In each state of the Markov chain, the first index denotes the backoff level, and the second index denotes the countdown value,
i.e., how many idle-channel slots must be observed before the frame is allowed to transmit.
The maximum backoff window at level $i$ is $W_i$, which is twice $W_{i-1}$.
In the top two rows, $W_0$ is a QoS parameter denoting the minimum contention window.
A smaller $W_0$ means that the STA/AC competes more aggressively for channel access.
After the $i^{\rm th}$ collision, a random integer is drawn uniformly from $0$ to $W_i-1$, which decides the countdown
before the next transmission attempt.

As we can see, this Markov chain follows the behavior described in IEEE~802.11e~\cite{80211e}.
Each AC of each STA runs through these states before its frame becomes eligible for transmission.
Each STA/AC has its own frame availability probability (or normalized offered load) $q_l$ and attempt probability $\tau_l$.
However, the collision probability $p_l$ is typically very similar across STA/ACs, especially when the number of STA/ACs is large.
In other words, each STA/AC executes its own backoff state machine, while the ``crowdedness'' of the wireless channel is implicitly
coupled across nodes through the mutual collision probability.

In Figure~\ref{fig:MarkovChart}, $P$ denotes the probability that the wireless channel is idle.
The terms shown in blue are the transition probabilities after a frame transmission from states other than $(0,0)_e$.
After such a transmission, the process enters state $(0,0)$ with probability $q$ if there is a frame waiting to be transmitted,
and enters state $(0,0)_e$ with probability $1-q$ if there is no frame waiting.
The terms shown in red correspond to the four possible transitions out of state $(0,0)_e$:
\begin{enumerate}
    \item With probability $1-q$, there is no data frame to transmit and the process stays in state $(0,0)_e$.
    \item With probability $q(1-P)$, at least one data frame is waiting, but the channel is not idle, and the process enters one of the states in the second row.
    \item With probability $qP(1-p)$, at least one data frame is waiting and it is transmitted successfully when the channel is idle (no collision); the process enters one of the states in the top row.
    \item With probability $qPp$, the same as the previous case except that a collision occurs; the process enters one of the states in the second row.
\end{enumerate}
Note that the events of (1) a frame becoming ready, (2) sensing an idle channel, and (3) encountering a collision (or not) occur sequentially
immediately after the system is in state $(0,0)_e$.

We next list three equations for each STA/AC: the equations in $q_l$, $p_l$, and $\tau_l$.

\paragraph{1) Frame availability probability $q_l$.}
The probability of frame availability (normalized offered load) is
\begin{equation}
    q_l = 1 - \exp(-\lambda_l E_S),
\end{equation}
where we assume a Poisson process with mean arrival rate $\lambda_l$, which is the reciprocal of the inter-arrival time
modeled in Section~\ref{Sec:TimeStructure}.  The term $E_S$ is the average time duration spent per state transition in the Markov chain,
averaged over three situations: (i) no transmission, (ii) one successful transmission, and (iii) collisions:
\begin{equation} \label{eq:ES}
    E_S = (1 - P_{\rm tr}) \sigma
    + \sum^n_{i = 1} P_{S_i} T_{S_i}
    + \sum_{r=2}^n \sum_{1 \le k_1 < \ldots < k_r \le n} P_{c,k_1, \ldots, k_r} T_{c,k_1, \ldots, k_r}.
\end{equation}
Here, the average durations of the three cases are respectively the slot time $\sigma$, the successful transmission time for the
$i^{\rm th}$ STA/AC denoted by $T_{S_i}$, and the average recovery time from a collision among STA/ACs indexed by
$k_1,\ldots,k_r$ denoted by $T_{c,k_1,\ldots,k_r}$.  These durations are quantities we expect to collect from the network
and are therefore treated as known to the DT.  On the other hand, the probabilities of the three cases are functions of the
attempt probabilities $\{\tau_i\}$ and are expressed as
\begin{align}
    P_{\rm tr} &= 1 - \prod_{i=1}^n (1 - \tau_i), \\
    P_{S_i} &= \tau_i \prod_{j \ne i} (1 - \tau_j), \\
    P_{c,k_1, \ldots, k_r} &= \prod_{i=1}^r \tau_{k_i} \, \prod_{j \ne k_1, k_2, \ldots, k_r} (1 - \tau_j).
\end{align}

\paragraph{2) Attempt probability $\tau_l$.}
The attempt probability can be obtained from the steady-state distribution of the Markov chain.
A transmission attempt occurs when the process is in (i) state $(0,0)_e$ and a frame is ready while the channel is idle,
or (ii) one of the states $(i,0)$ for $i\ge 0$ and a frame is ready while the channel is idle after $i$ collisions.
After algebraic manipulation (with the STA/AC subscript omitted for clarity), we obtain
\begin{equation}
    \tau = b(0,0)_e \left(
    \frac{q^2 W_0}{(1-p)(1-q)\left(1-(1-q)^{W_0}\right)}
    - \frac{q^2 P_{\rm idle}}{1-q}
    \right),
\end{equation}
where the steady-state probability of state $(0,0)_e$ is
\begin{align}
    \frac{1}{b(0, 0)_e} &=
    (1-q) + \frac{q^2 W_0 (W_0 + 1)}{2\left(1-(1-q)^{W_0}\right)}  \nonumber\\
    &\quad + \frac{q(W_0+1)}{2(1-q)}
    \left(
    \frac{q^2 W_0}{1 - (1-q)^{W_0}} + (1 - P_{\rm idle})(1 - q) - q P_{\rm idle} (1-p)
    \right) \nonumber\\
    &\quad + \frac{p q^2}{2(1 - q)(1 - p)}
    \left(
    \frac{W_0}{1 - (1 - q)^{W_0}} - (1-p) P_{\rm idle}
    \right)
    \left(
    2 W_0 \frac{1 - p - p(2p)^{m-1}}{1 - 2p} + 1
    \right),
\end{align}
and the probability that the $l^{\rm th}$ STA/AC observes an idle channel is
\begin{equation}
    P_{{\rm idle}, l} = \prod_{i \ne l}^n (1 - \tau_i) = 1 - p_l,
\end{equation}
which corresponds to $P$ in Figure~\ref{fig:MarkovChart}.

\paragraph{3) Collision probability $p_l$.}
The third nonlinear equation is
\begin{equation}
    1 - p_l = \prod_{j \neq l} (1 - \tau_j),
\end{equation}
for $l = 1, \ldots, n$.

By solving the system of nonlinear equations above, we typically obtain a unique and sensible solution set
$\{q_l\}$, $\{\tau_l\}$, and $\{p_l\}$.  We can then compute the normalized throughput $S$ as
\begin{equation}
    S = \frac{P_S P_{\rm tr} E[P]}{(1 - P_{\rm tr}) \sigma + P_S P_{\rm tr} T_S + P_{\rm tr}(1 - P_S) T_C},
\end{equation}
where $P_S$ is defined as
\begin{equation}
    P_S = \frac{\sum_i P_{S_i}}{P_{\rm tr}},
\end{equation}
and the average payload duration $E[P]$ is information collected from the network.

In addition, the average delay $E[D]$ can be approximated as
\begin{equation}
    E[D] = E \left[ \sum_{i = 1}^N L_i \right] = E[N] E[L],
\end{equation}
where $E[L]$ (equal to $E_S$ in (\ref{eq:ES})) is the average time per state, and $E[N]$ is the average number of states
visited before the next successful transmission.  After further manipulation, we obtain
\begin{equation}
    E[N] = \frac{W_0}{2} \frac{1 - (2p)^{m+1}}{1 - 2p}
    + \frac{W_0 p (2p)^m}{2} \frac{1}{1 - p}
    + \frac{1}{2} \frac{1}{1 - p}.
\end{equation}


\begin{figure}
    \centering
    \includegraphics[width=1\linewidth]{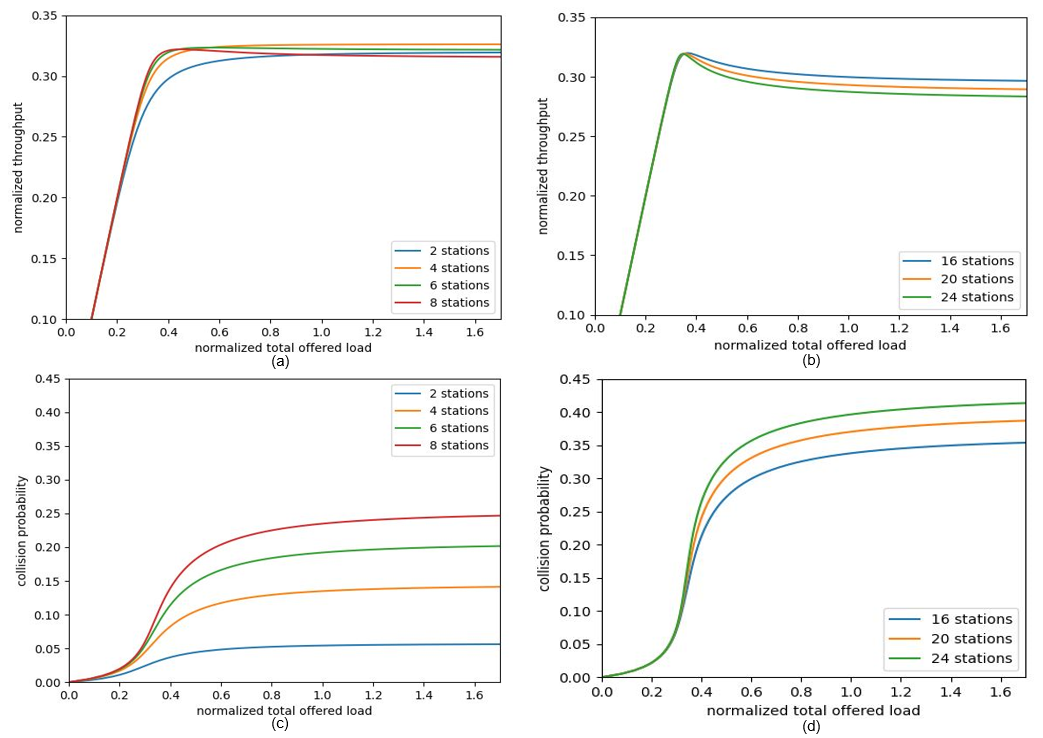}
    \caption{(a), (b) Normalized throughput and (c), (d) collision probability vs normalized total offered load 
    with single-rate single-QoS non-saturated traffic}
    \label{fig:NonSaturationThroughput}
\end{figure}

\begin{figure}
    \centering
    \includegraphics[width=1\linewidth]{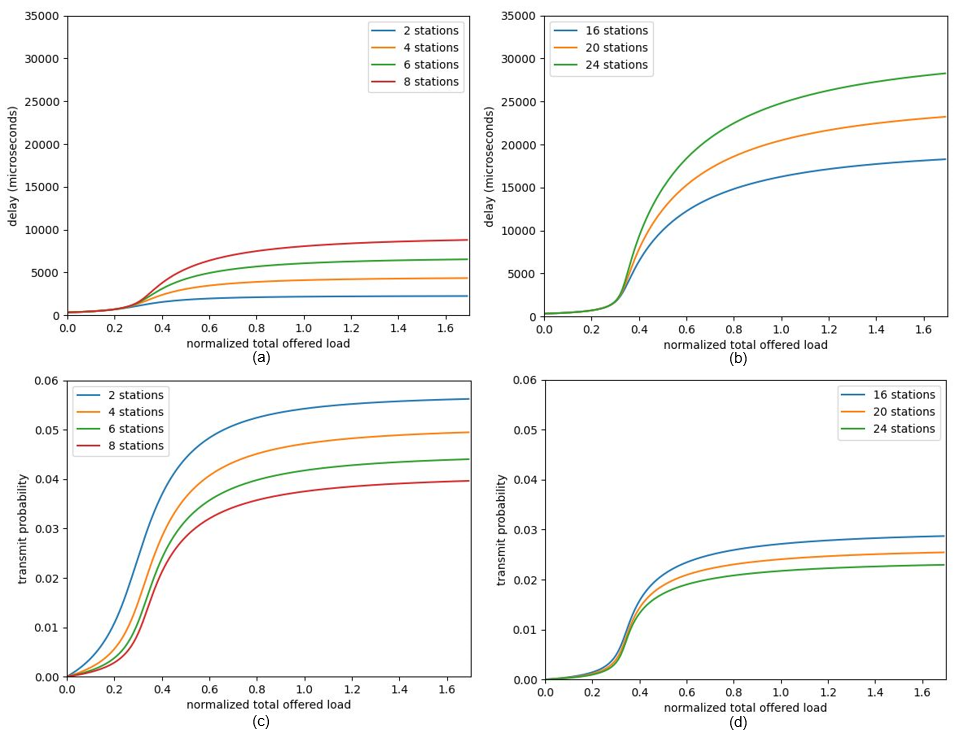}
    \caption{(a), (b) Delay and (c), (d) transmit probability vs normalized total offered load 
    with single-rate single-QoS non-saturated traffic}
    \label{fig:NonSaturationDelay}
\end{figure}

We want the cost functions to be efficient to evaluate so that we can scan a large number of scenarios for risk detection.
However, solving a system of $3K$ nonlinear equations may still be too expensive.  Next, we simplify this computation in three stages.

First, we verify the accuracy of the formulation above, which will be used to numerically evaluate the key performance measures
of a CSMA-CA network.
Second, we simplify the computation for a single-QoS multi-rate network.  In this setting, all participating STAs use the same QoS
parameters (e.g., $W_0$, $m$, AIFSN, and TXOP) but may have different normalized offered loads.  This is common because each STA may be
at a different distance from its associated AP, and therefore may be assigned a different MCS.  Even if the requested throughput is the same,
different MCS values lead to different channel occupancy time, and hence different normalized offered loads.
Third, we take a further step to simplify the computation for a multi-QoS and multi-rate network.

An important task at this point is to ensure the accuracy of the analytical results above before using them as cost functions in both the risk monitor and the risk mitigator.
It is tedious and time-consuming to run system-level simulations for a wireless network with a few APs and tens of STAs over
a long time horizon (e.g., weeks).  Fortunately, there are published works (e.g., \cite{Malone2007}) that use the ns-2
(network simulator-2) to evaluate and summarize results for various scenarios.  By applying their scenarios to the formulation above
and comparing with their reported results, we can assess the accuracy of our analytical formulation.

By applying the simulation scenarios in \cite{Malone2007} to the formulation above, we obtain the normalized throughput,
collision probability, frame delay, and transmission attempt probability shown in Figures~\ref{fig:NonSaturationThroughput}
and~\ref{fig:NonSaturationDelay}.
Although these figures do not explicitly overlay the ns-2 results, we have cross-checked our computed curves against the ns-2
simulation results reported in \cite{Malone2007}, and the outcomes are extremely close.

We now move on to further simplify the computation in a way that closely approximates the analytical results while greatly reducing
evaluation time, enabling efficient scenario scanning.

\subsubsection{Single-QoS Multi-Rate Networks}

For single-QoS scenarios, by examining the single-rate results shown in
Figures~\ref{fig:NonSaturationThroughput}--\ref{fig:NonSaturationDelay}, we make the following observations:
\begin{enumerate}
    \item In Parts (a) and (b) of Figure~\ref{fig:NonSaturationThroughput}, the normalized throughput is approximately linear at low offered load.
    This implies that the network can fulfill traffic demands from a wide range of STA populations with low collision probability, as shown in
    Parts (c) and (d) of Figure~\ref{fig:NonSaturationThroughput}, negligible frame delay as shown in
    Parts (a) and (b) of Figure~\ref{fig:NonSaturationDelay}, and reasonably low transmission attempt probability for each STA as shown in
    Parts (c) and (d) of Figure~\ref{fig:NonSaturationDelay}.
    \item From Parts (a) and (b) of Figure~\ref{fig:NonSaturationThroughput}, as the normalized offered load increases, the normalized throughput peaks
    at roughly the same offered load value, except for scenarios with a very small number of STAs (e.g., 2 or 4 STAs).  This point coincides with the onset of a
    significant rise in collision probability, as shown in Parts (c) and (d) of Figure~\ref{fig:NonSaturationThroughput}, and with a significant rise in
    delay and transmission attempt probability, as shown in Figure~\ref{fig:NonSaturationDelay}.
    \item An ideal operation point for network optimization should be slightly below the throughput peak, so that the network can absorb traffic variations
    over time and remain robust.  We identify this neighborhood as the \emph{latency performance upper bound}, and recommend adding an additional
    margin to accommodate implementation penalties.  To avoid confusion, this ``latency upper bound'' is the point beyond which frame delay and collision
    probability increase rapidly and throughput begins to saturate or decline.  We therefore operate with a margin below this bound.
    \item When the total offered load increases toward the latency upper bound, the normalized throughput reaches its peak and begins to saturate while the
    frame delay increases sharply.  It is often easier to monitor the normalized throughput than the frame delay because the throughput transition is smoother.
    More importantly, before saturation the overall normalized throughput closely tracks the overall offered load with a slope close to one.  Therefore, for the rest
    of this paper we refer to the latency upper bound interchangeably in terms of either the normalized offered load or the normalized throughput.
    \item It is clear from Parts (a) and (b) of Figure~\ref{fig:NonSaturationThroughput} that for network traffic with the same QoS settings generated by more than
    two STAs, the performance curves are very similar as long as the \emph{overall} offered load is the same.  This is especially true before throughput saturation,
    when collision probability is low and there is little interaction among traffic streams.  From the viewpoint of a single STA, what matters is the aggregate
    contention level reflected by the collision probability, not the exact number of contenders.  Hence, to estimate the latency upper bound, we can focus on
    the overall normalized offered load (the sum over all STAs) rather than the detailed distribution across STAs.  This observation will also be shown to hold for
    multi-rate scenarios.
\end{enumerate}

\paragraph{Complexity reduction for single-QoS multi-rate networks.}
We now exploit the above conclusion to simplify the computation of latency upper bounds.
Consider a network where all $K$ STAs apply the same QoS settings.  In practice, each STA will typically have a distinct normalized offered load due to both
channel-rate differences (different RCPI and thus different MCS) and heterogeneous traffic patterns.  In principle, computing delay, collision probability,
attempt probability, and throughput for this $K$-rate system requires solving a $3K$-unknown nonlinear system for each DT-emitted scenario.
This would be prohibitive for (i) a risk monitor that scans many scenarios, and (ii) a risk mitigator that evaluates many candidate configurations during
gradient/exhaustive search.

Based on the empirical and analytical observation that the latency upper bound is primarily determined by the \emph{overall} normalized offered load under a fixed
QoS setting, we approximate the multi-rate system by an equivalent single-rate system to determine the upper bound.  Concretely:
\begin{enumerate}
    \item For each QoS setting, solve a single-rate model (a system of only three nonlinear equations) to obtain the latency upper bound in terms of the
    overall normalized offered load (or equivalently the saturation throughput point).
    \item For any DT-emitted scenario with $K$ heterogeneous STAs (i.e., a $K$-rate system), the risk monitor only needs to sum the normalized offered loads of all
    $K$ STAs and compare the sum against the pre-computed upper bound (with a safety margin).
    \item If the summed offered load approaches or exceeds the upper bound within a prescribed margin, the risk monitor triggers an alarm and the risk mitigator
    searches for a better configuration.
    \item During the risk mitigator's search, for each candidate configuration we again compute (or look up) the single-rate upper bound corresponding to the
    QoS setting, and evaluate the \emph{margin} between the upper bound and the summed offered load of the candidate scenario.
    The configuration that maximizes this margin is chosen as the mitigation solution.
\end{enumerate}
With this approximation, the dominant computation in both the risk monitor and the risk mitigator is reduced from repeatedly solving $3K$ nonlinear equations to
solving only three equations per QoS setting, plus simple summations across STAs.

However, this conclusion does not directly extend to scenarios where different STAs (or different ACs) adopt different QoS settings, because the mapping from
aggregate offered load to collision probability and delay depends on the QoS parameters (e.g., $W_0$, $m$, AIFSN, TXOP) in a class-dependent manner.
We therefore next examine multi-QoS scenarios and develop a comparable complexity-reduction strategy for both the risk monitor and the risk mitigator.

\subsubsection{Multi-QoS Multi-Rate Networks}

In a realistic Wi-Fi network, the traffic is not only multi-rate as discussed earlier, but also inherently multi-QoS.
In IEEE~802.11e, four Access Categories (ACs) are typically supported: VO (voice), VI (video), BE (best effort), and BK (background).
Different ACs are recommended to use different combinations of QoS parameters, namely:
1) minimum contention window ${\rm CW}_{\rm min}$ (denoted as $W_0$),
2) maximum backoff level $m$,
3) arbitration inter-frame space number (AIFSN), and
4) allowable transmission opportunity (TXOP).
These parameters jointly change the aggressiveness of channel access and the effective payload efficiency, and therefore alter
the latency upper bound and the saturation throughput.

Consider a multi-QoS network with $K$ QoS groups.  Group $k$ has $n_k$ STAs and a fixed parameter set
$(W_{0,k}, m_k, {\rm AIFSN}_k, {\rm TXOP}_k)$.
If only group $k$ exists (all STAs in the network use the same QoS set $k$) and the offered load is high enough to reach saturation,
we denote the resulting \emph{fair} saturation throughput by $S_k$ (overall normalized throughput at saturation for that single-QoS network).
A simple and practically useful hypothesis is that the overall saturation throughput of the mixed network can be approximated by
a station-count-weighted average of the single-QoS saturation throughputs:
\begin{equation}\label{eq:ArithmaticSum}
    S \approx \frac{\sum_k n_k S_k}{\sum_k n_k}.
\end{equation}
This hypothesis suggests the following procedure: pre-compute $S_k$ for each QoS set $k$ (by solving the single-rate, single-QoS system),
then combine them using (\ref{eq:ArithmaticSum}) to obtain an overall saturation reference $S$ (or equivalently, an offered-load upper bound)
for fast risk scanning and fast mitigation search.
In fact, we justified this approximation and also the approximations made in the previous sub-section for single-QoS multi-rate, through numerical examples and find both satisfactory.
Interested readers can refer to author's YouTube channel \cite{HandsOnDigitalTwins}, where detailed 
stories are elaborated. 

\paragraph{Implications for fast upper-bound evaluation.}
With the full analytical formulation, multi-QoS multi-rate evaluation for up to roughly 30--50 stations is feasible,
but repeatedly solving a $3K$-unknown nonlinear system is still too costly for
(i) scanning many DT-emitted ``parallel-universe'' scenarios in the risk monitor, and
(ii) evaluating many candidate configurations during gradient search in the risk mitigator.

We therefore combine two layers of upper-bound screening:
\begin{itemize}
    \item \textbf{Layer~1 (Shannon throughput bound):} For each link, Shannon capacity provides a power-limited upper bound on achievable data rate,
    independent of CSMA/CA contention.  This bound is crucial for any system (Wi-Fi or cellular), and is especially dominant in scheduled systems
    (e.g., 5G/6G) where contention-induced delay is not the primary limiter.
    \item \textbf{Layer~2 (CSMA/CA latency bound via normalized throughput upper bound):} For Wi-Fi, contention creates a latency upper bound that manifests as a
    saturation/corner point in overall normalized throughput.  Normalized throughput is defined as the percentage of time used for successful, collision-free
    payload transmission over the available channel time, and it depends on contention dynamics and overhead in addition to receive power.
\end{itemize}
For multi-QoS coexistence, we approximate the mixed-network saturation behavior by pre-computing per-QoS saturation references and combining them using the
weighted-average rule in (\ref{eq:ArithmaticSum}).  This yields a low-complexity, scenario-scannable upper bound analogous to the single-QoS multi-rate case.

\paragraph{Summary procedure for computing the performance upper bound.}
We summarize a practical procedure that is efficient enough for both risk scanning and mitigation search:
\begin{enumerate}
    \item For a target AP/band, pre-compute normalized-throughput (latency) upper bounds for each of the four ACs
    using the QoS parameters announced by the AP for that AC, regardless of data rate (single-rate reference per AC).
    Denote the corresponding single-QoS saturation throughputs as $\{S_k\}$ (or the equivalent offered-load critical values).
    \item For each STA/AC, compute its offered load by summing both uplink and downlink demands (reciprocal channels share the same contention pool).
    \item Use receive power (or SINR) to determine the best-allowed MCS via Shannon bound with an implementation margin,
    and convert each STA/AC throughput demand into \emph{normalized offered load} by accounting for payload efficiency, MAC/PHY headers, and mandatory gaps.
    If average frame size is unavailable, assume MTU payload to obtain an upper bound.
    \item Repeat the previous steps per AC and compute an effective per-AC aggregate normalized offered load.
    Use the per-AC saturation references to form an overall latency upper bound, e.g., by combining the per-AC saturation throughputs through a weighted rule
    analogous to (\ref{eq:ArithmaticSum}) using the number of contending STA/AC entities in each AC.
    \item Sum the normalized offered loads of all STA/AC contenders in the OBSS and compare the sum against the computed upper bound (with a pre-defined safety
    margin).  If the sum is below the bound, the difference is the latency margin for that AP/band.
    \item If any AP/band has a margin smaller than a threshold, trigger an alarm and let the risk mitigator search for a configuration that maximizes the minimum
    margin across all AP/bands.
\end{enumerate}

\section{\label{sec:Simulations}Simulations}

\begin{figure}
    \centering
    \includegraphics[width=1\linewidth]{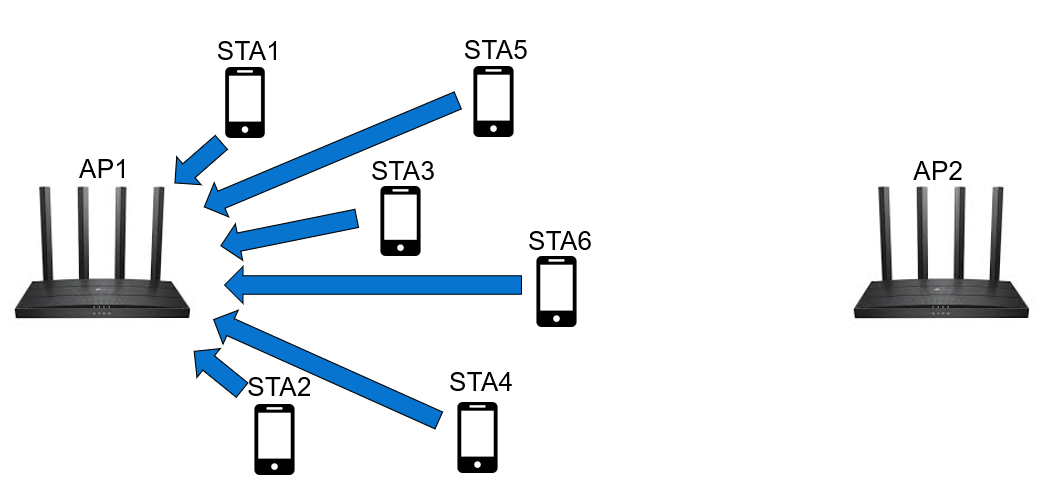}
    \caption{A network performance degradation scenario that requires load balancing.}
    \label{fig:TC1Scenario}
\end{figure}

\begin{figure}
    \centering
    \includegraphics[width=0.5\linewidth]{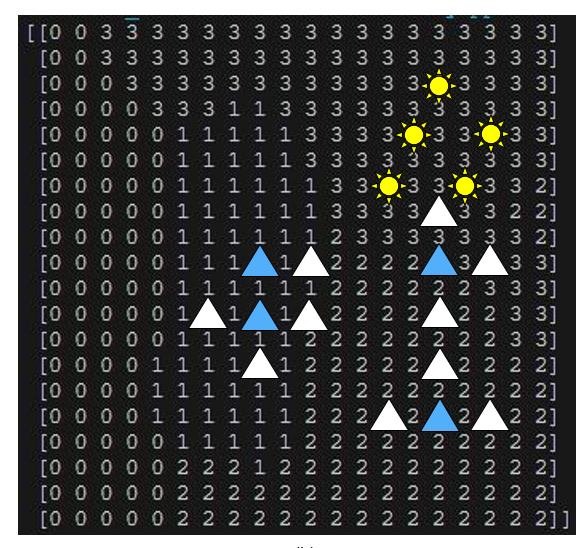}
    \caption{STA locations in the simulation scenarios.}
    \label{fig:StaLocation}
\end{figure}

\begin{figure}
    \centering
    \includegraphics[width=1\linewidth]{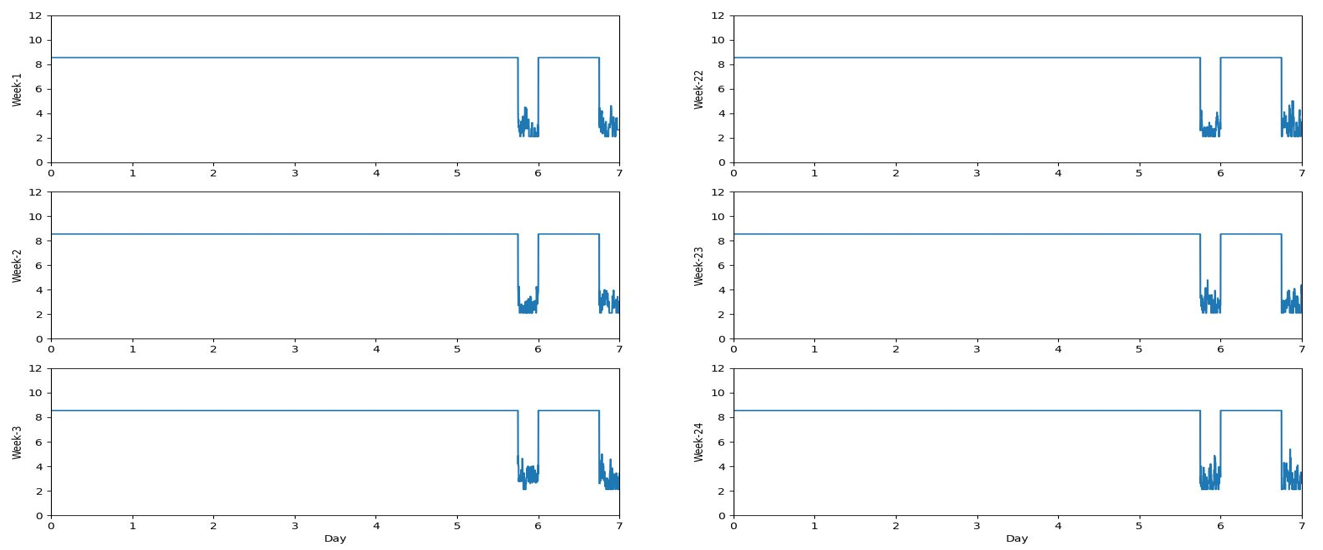}
    \caption{Average distance from STA to AP\#3, averaged over all STAs.}
    \label{fig:TC1Distance}
\end{figure}

\begin{figure}
    \centering
    \includegraphics[width=1\linewidth]{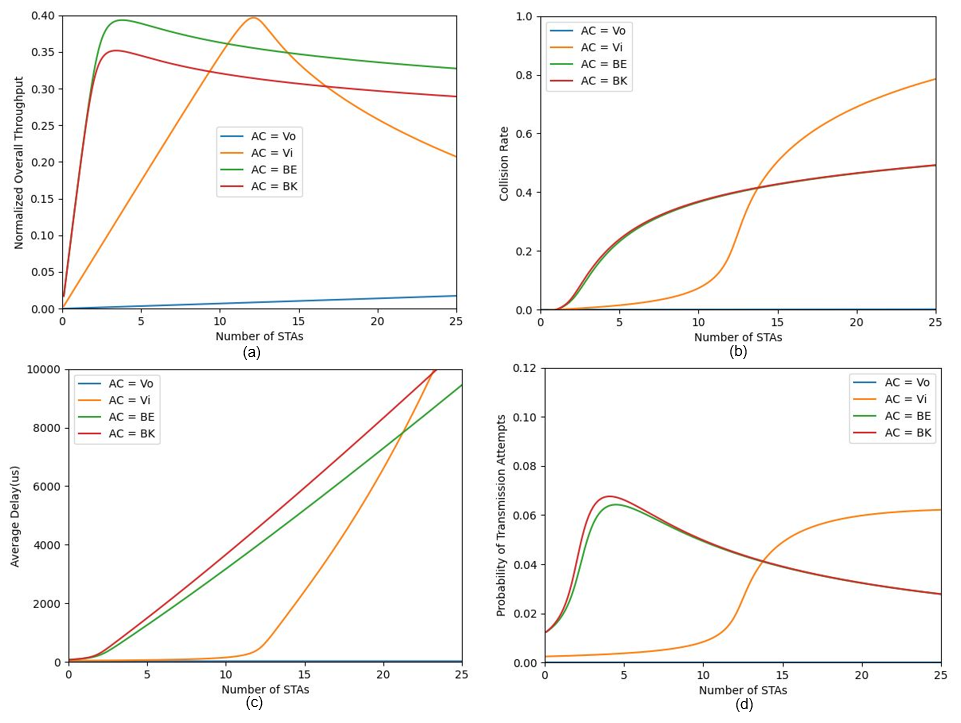}
    \caption{Performance measures in calculating the upper bounds for each of the four ACs:
    (a) overall normalized throughput, (b) collision probability, (c) average delay, and
    (d) probability of transmission attempts.}
    \label{fig:TC1UpperBound}
\end{figure}

Before presenting the simulation results, we emphasize that the simulations in this section are intended primarily as a fast and controlled means to validate the proposed predictive and preventive framework, rather than as a substitute for real-world Digital Twin deployment.  Simulation allows systematic exploration of time-dependent traffic evolution and network reconfiguration under repeatable conditions, which is essential for isolating and verifying the core algorithmic behavior.

A fully operational Digital Twin ultimately requires continuous interaction with a Physical Twin through real measurements and control actions.  Such real-world experimental validation is currently under active development by the authors, and will be reported separately once sufficient long-term measurement data and system integration are completed.

To demonstrate that the proposed algorithm works, we construct wireless scenarios that are likely to experience
performance degradation and verify that these problems can be identified and mitigated by the proposed framework.
Because the proposed algorithm uses throughput bounds and frame-latency bounds as cost functions, many practical scenarios
naturally fit its scope: there is little reason for users to complain if both the achievable throughput (in the sense of
power-limited capacity) and the contention-induced latency (in the sense of CSMA-CA saturation) are properly addressed.

\subsection{Simulation Scenarios}

As shown in Figure~\ref{fig:TC1Scenario}, we consider a representative performance-degradation scenario where load balancing is required.
In this example, the proposed algorithm does not merely count how many STAs are associated with the overloaded AP (AP\#1), nor does it
only consider the total offered load; rather, it also accounts for receive power and the corresponding Shannon capacity at each STA.
For simplicity, assume all STAs request the same offered load.

If we only count the number of STAs associated with AP\#1, we might arbitrarily choose STA\#1, STA\#2, and STA\#3 to re-associate with a
neighboring AP (AP\#2) to relieve AP\#1.  This is clearly suboptimal, because STA\#4, STA\#5, and STA\#6 are closer to AP\#2 and thus
are cheaper to serve from AP\#2, while being more costly for AP\#1.  However, even moving all three boundary STAs (STA\#4--STA\#6) may
still be infeasible: these STAs are near the cell boundary, and therefore consume substantially more wireless resources due to lower RCPI
(and potentially higher interference), while STA\#1--STA\#3 cost very little to serve.  It is therefore possible that AP\#2 does not have
enough remaining resources to simultaneously serve STA\#4, STA\#5, and STA\#6.  In contrast, the proposed algorithm may recommend a more
selective reassociation (e.g., moving only STA\#5 and STA\#6), after jointly considering the resource margins of both AP\#1 and AP\#2.
This yields a feasible and robust configuration in which both the Shannon-throughput feasibility and the latency upper bound in terms of
normalized offered load are satisfied.

To construct a time-varying test case that repeatedly triggers load imbalance, we extend the time-structure example described in
Section~\ref{Sec:TimeStructure}: People in the household
tend to gather in the living room on weekend nights.  The household layout and the locations of four APs are given in
Figure~\ref{fig:HouseLayout}, where the living room is located at the top-right corner.  To keep the test case focused on load balancing,
we assume the four APs operate on non-overlapping channels (hence no mutual co-channel interference), and the backhaul among APs is wired
Ethernet (hence no wireless backhaul traffic competes for the same channel resources).

We further assume that all STAs are stationary and appear at the locations shown in Figure~\ref{fig:StaLocation}.  In that figure,
triangular markers indicate STA locations during weekdays and non-weekend nights; blue triangles represent permanent residents, while
white triangles represent visitors.  On weekend nights, yellow sun-shaped markers indicate the STA locations after the crowd converges
to the living room region.

We generate traffic traces over a total duration of 24 weeks.  To verify that the generated traffic pattern indeed matches the intended
load-imbalance behavior, we plot the average distance between STAs and AP\#3 (the overloaded AP on weekend nights), averaged over all
STAs, as shown in Figure~\ref{fig:TC1Distance}.  The plot includes the first three and the last three weeks in the simulation period.
The pattern is consistent across weeks: the average distance to AP\#3 is significantly shorter on weekend nights, confirming that the
scenario repeatedly concentrates STA presence near AP\#3 and thus stresses its resource margin.

Based on the traffic model in Section~\ref{Sec:TimeStructure}, with 14 STAs each generating four AC traffic demands in both uplink and
downlink, and based on typical IEEE~802.11ax parameters~\cite{80211ax}, we next compute the upper bound of each AC as summarized at
the end of Section~\ref{sec:LatencyBounds}.  The results are shown in Figure~\ref{fig:TC1UpperBound}.  We assume an operating bandwidth
of 20~MHz with one spatial stream.  The obtained bounds can be scaled proportionally; for example, an 80~MHz bandwidth with 8 spatial
streams provides approximately $32\times$ the upper bounds shown here.

We assume a target throughput demand of 100~kbps per STA for VO traffic and 5~Mbps per STA for VI traffic.  For BE and BK traffic, we
assume a minimum acceptable download rate of 25~Mbps per STA.  Several observations can be made from
Figure~\ref{fig:TC1UpperBound}.  First, VO traffic requires on the order of hundreds of full-rate STAs (about 570 in this configuration)
to approach the saturation turning point, and thus VO can be ignored in this particular stress test.  Second, for each AC, the turning
point (where throughput saturates and collision/delay/attempt probability start to rise sharply) coincides across the four subfigures,
confirming that normalized throughput saturation is a reliable indicator of the latency upper bound.  Finally, VI traffic is typically
assigned more aggressive EDCA parameters than BE and BK per IEEE~802.11e, so VI can enjoy an advantage when traffic is heavy but not
extreme.  However, when traffic becomes excessively heavy, this aggressiveness causes VI to pound the channel more intensely, leading to
worse overall behavior not only in normalized throughput but also in collision probability, average frame delay, and transmission-attempt
probability.

Readers interested in implementation details can refer to author's YouTube channel \cite{HandsOnDigitalTwins}, where detailed stories are elaborated. 

\subsection{Simulation Results}

\begin{figure}
    \centering
    \includegraphics[width=1\linewidth]{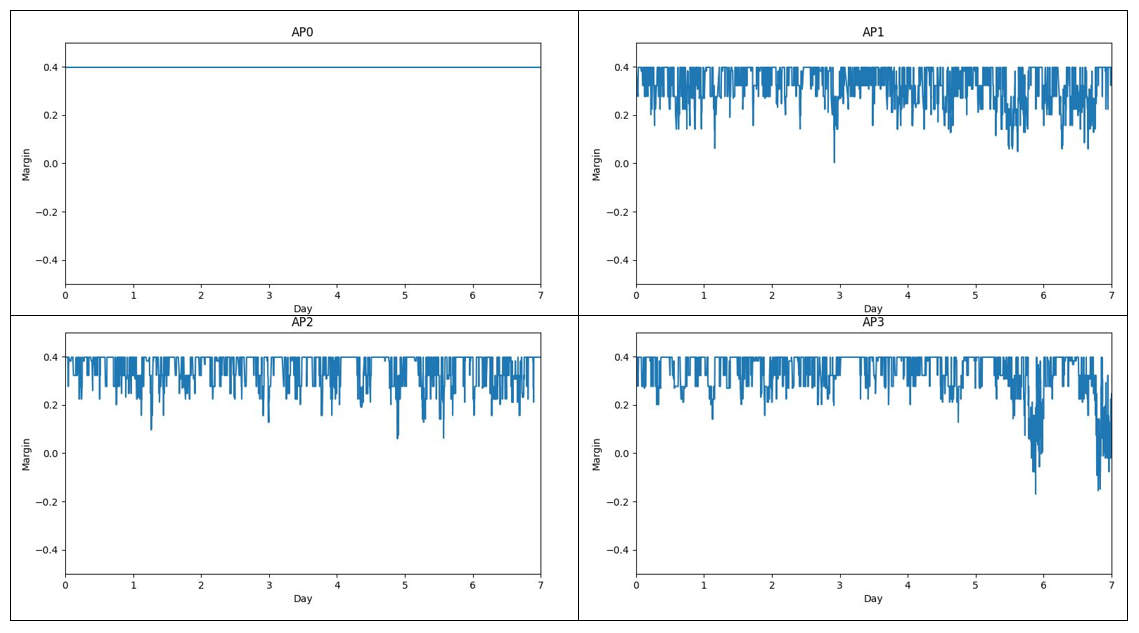}
    \caption{Performance margin on the simulated PT scenario over a week, for each of the 4 APs in the service area.}
    \label{fig:TC1MarginBefore}
\end{figure}

\begin{figure}
    \centering
    \includegraphics[width=1\linewidth]{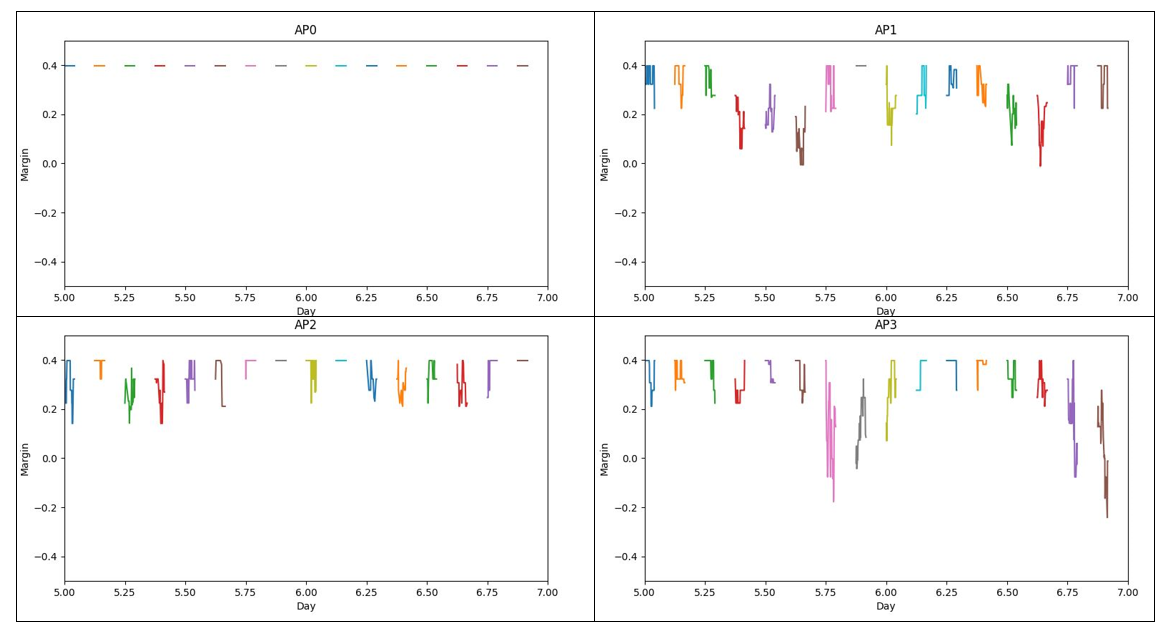}
    \caption{Performance margin calculated based on scenarios predicted by the DT, for each of the 4 APs in the service area.
    The DT periodically predicts network scenarios one hour into the future, with predictions generated every 3 hours.}
    \label{fig:TC1MarginDT}
\end{figure}

The proposed algorithm is applied to compute the performance margin for each AP, where the minimum margin indicates the most urgent risk to mitigate
if it falls below a predefined acceptable threshold.  We evaluate the resource margin using the latency upper bound, based on PT data, for each AP
with 20~MHz channel bandwidth and two spatial streams.  Figures~\ref{fig:TC1MarginBefore} and \ref{fig:TC1MarginDT} show the resulting margins for
each of the four APs in the service area, computed respectively from the simulated PT scenario and from the DT-predicted scenarios.

Here, the margin is defined as the leftover quota for an OBSS to support additional offered load contributed by either the AP or any STA within this
OBSS.  When the margin drops below zero, it indicates that the OBSS cannot support the offered-load requests of the AP and associated STAs, even under
ideal traffic shaping and optimal scheduling.

Figure~\ref{fig:TC1MarginBefore} shows that AP\#3 is overloaded on both weekend nights in the PT scenario, while the other three APs remain feasible.
No traffic is assigned to AP\#0.  The margin at AP\#1 is tighter because two STAs corresponding to permanent residents share AP\#1's resource; in
principle, additional optimization is possible.  Figure~\ref{fig:TC1MarginDT} shows that the DT predicts exactly the same events---namely, the
overload of AP\#3 on both weekend nights.

\begin{figure}
    \centering
    \includegraphics[width=1\linewidth]{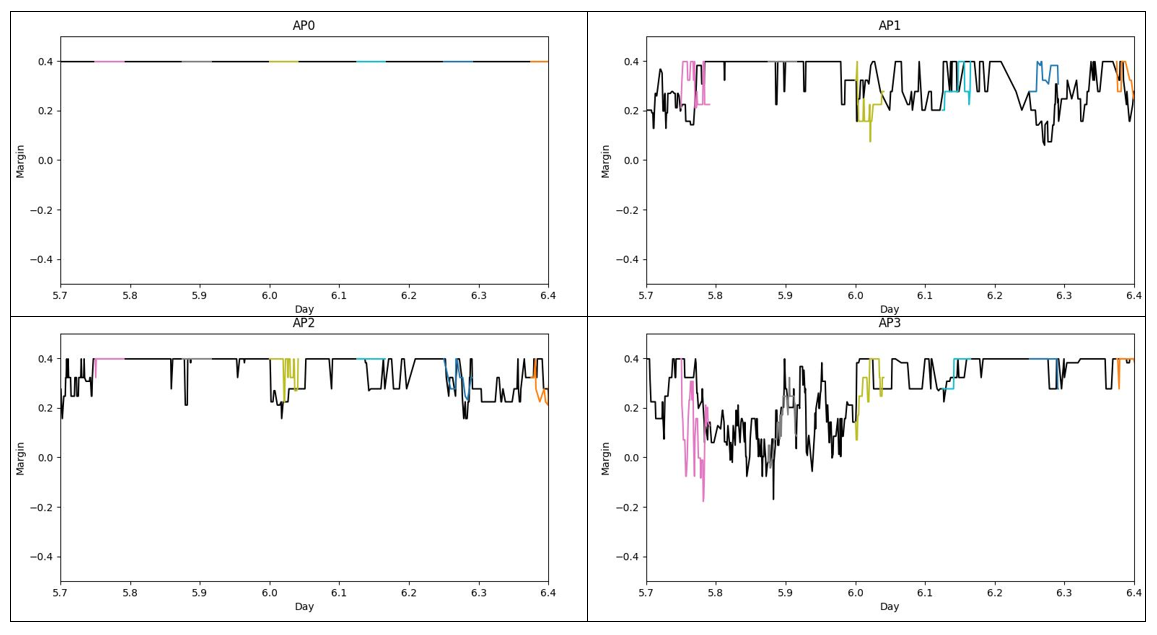}
    \caption{Overlapping of Figures \ref{fig:TC1MarginBefore} and \ref{fig:TC1MarginDT}, after zooming in around Saturday night.}
    \label{fig:TC1MarginRoomIn}
\end{figure}

To examine timing more closely, Figures~\ref{fig:TC1MarginBefore} and \ref{fig:TC1MarginDT} are temporally aligned and zoomed in around Saturday
night, as shown in Figure~\ref{fig:TC1MarginRoomIn}.  We observe that the risk monitor triggers alarms based on DT-predicted scenarios roughly one to
two hours before performance degradation actually occurs in PT.  This lead time arises because DT prediction is driven by historical statistics, while
the realized PT evolution may deviate from those statistics; therefore, both false alarms and missed detections are possible.  False alarms are not
necessarily harmful: even if degradation does not occur in a particular instance, it is still beneficial to ensure the network can handle the event
should it occur.  To reduce missed detections, one may increase the number of DT scenario samples (i.e., the number of predicted parallel universes),
or increase alarm sensitivity by raising the minimum acceptable margin threshold.

\begin{figure}
    \centering
    \includegraphics[width=1\linewidth]{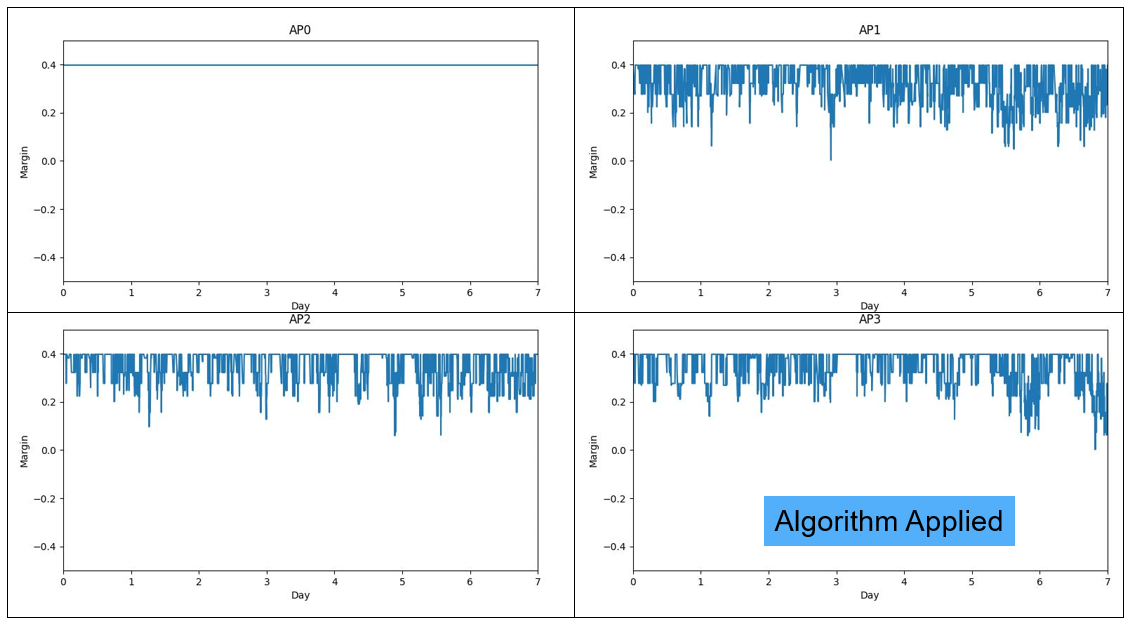}
    \caption{Performance margin calculated after the proposed optimization is applied to the PT scenario, for each of the 4 APs in the service area.}
    \label{fig:TC1MarginAfter}
\end{figure}

Once the risk monitor flags low-margin DT scenarios, the risk mitigator reconfigures the network to avoid these alarm-triggering outcomes.  After the
proposed optimization is applied, Figure~\ref{fig:TC1MarginAfter} shows that the negative margins previously observed at AP\#3 disappear.  A closer
inspection reveals that some STAs served by AP\#3 on weekend nights are re-associated to either AP\#1 or AP\#2, consistent with the margin trade-off:
the times when AP\#3's margin increases coincide with decreases in the margin of AP\#1 and/or AP\#2, as shown in Figure~\ref{fig:TC1MarginAfter}.

\section{Conclusions}
This paper presents a predictive and preventive framework for managing performance degradation in Wi-Fi networks using a Digital Twin.
By combining DT-based scenario generation with analytically derived performance upper bounds, the proposed approach enables early detection of risky operating conditions and proactive mitigation before user experience deteriorates.

Unlike conventional reactive optimization methods, the proposed framework explicitly accounts for both physical-layer constraints and MAC-layer contention behavior.
The Shannon capacity bound captures the limitations imposed by wireless channel conditions, while the normalized throughput–based latency bound characterizes the onset of instability caused by excessive contention under CSMA/CA.
Together, these bounds provide a principled and computationally efficient way to assess how close a network is to performance degradation.

A representative Wi-Fi mesh scenario was constructed to validate the effectiveness of the proposed approach.
Simulation results show that the Digital Twin can successfully predict future overload events with sufficient lead time, allowing the risk monitor to trigger alarms in advance.
By applying the same analytical bounds in the mitigation stage, the framework identifies feasible reconfiguration strategies—such as load balancing across APs—that restore performance margins without introducing unnecessary complexity.
This conservative design philosophy results in stable, explainable decisions that are well suited for real deployments.

Although the paper focuses on Wi-Fi networks operating under CSMA/CA, the methodology extends naturally to other wireless systems.
In scheduled systems such as 5G and 6G, where latency behavior is governed by different mechanisms, the Shannon capacity bound alone may suffice as the primary performance constraint.
Future work may further extend the framework to cover emerging features such as multi-link operation, advanced beamforming feedback, and tighter integration between physical-layer sensing and network control.

Overall, this work demonstrates that analytical performance bounds, when combined with predictive Digital Twins, offer a powerful tool for scalable, preventive network optimization.
We hope that this framework will encourage further research into proactive, model-driven management of complex wireless systems and inspire practical implementations in next-generation networks.

\bibliographystyle{IEEEtran}
\bibliography{references}

@ARTICLE{Malone2007,
  author    = {David Malone and Ken Duffy and Doug Leith},
  journal   = {IEEE/ACM Transactions on Networking},
  title     = {Modeling the {IEEE} 802.11 Distributed Coordination Function in Nonsaturated Heterogeneous Conditions},
  year      = {2007},
  volume    = {15},
  number    = {1},
  pages     = {159--172},
  doi       = {10.1109/TNET.2006.890136}
}

@TECHREPORT{3GPP-38901,
  author      = {{3GPP}},
  title       = {{TR} 38.901: Study on Channel Model for Frequencies from 0.5 to 100 {GHz}},
  institution = {3rd Generation Partnership Project},
  year        = {2023},
  url         = {https://www.3gpp.org/DynaReport/38901.htm}
}

@ARTICLE{80211e,
  author    = {{IEEE 802.11e}},
  journal   = {IEEE Std 802.11e-2005},
  title     = {Wireless {LAN} Medium Access Control ({MAC}) and Physical Layer ({PHY}) Specifications—{QoS} Enhancements},
  year      = {2005},
  doi       = {10.1109/IEEESTD.2005.97890}
}

@ARTICLE{80211ax,
  author    = {{IEEE 802.11ax}},
  journal   = {IEEE Std 802.11ax-2021},
  title     = {High Efficiency {WLAN}},
  year      = {2021},
  doi       = {10.1109/IEEESTD.2021.9442429}
}

@MISC{ns3website,
  author = {{ns-3 Consortium}},
  title  = {ns-3 Network Simulator},
  year   = {2024},
  url    = {https://www.nsnam.org/}
}

@ARTICLE{ParallelUniverses,
  author  = {A. McKenzie},
  journal = {Axiomathes},
  title   = {Reality and Super-Reality: Properties of a Mathematical Multiverse},
  year    = {2019},
  doi     = {10.1007/s10516-019-09466-7}
}

@MISC{HandsOnDigitalTwins,
  author = {J.-T. Chen},
  title  = {Hands-On Digital Twins (YouTube Channel)},
  year   = {2026},
  note   = {Available online}
}

@article{malone2011,
  author={Malone, David and Kenneally, Ed and Kelly, Mark and Grimes, Graham},
  title={Modelling the capacity of IEEE 802.11e EDCA under non-ideal conditions},
  journal={IEEE Transactions on Mobile Computing},
  volume={10},
  number={12},
  pages={1715--1726},
  year={2011},
  publisher={IEEE}
}

@article{fall1999,
  author={Fall, Kevin and Varadhan, Kannan},
  title={{ns} notes and documentation},
  year={1999},
  journal={UC Berkeley, LBL, USC/ISI},
  note={Available from: \url{www.isi.edu}}
}

@misc{viavi_ndt,
  author={{VIAVI Solutions}},
  title={Network Digital Twin: Accelerating 5G/6G Deployment and Optimization},
  howpublished={White Paper/Product Page},
  year={2024},
  note={Specific date from search result was September 2025, but a general year is more standard for BibTeX},
  url={https://www.viavisolutions.com/en-us/solutions/network-digital-twin}
}

@misc{arxiv_genai_dt,
  author={Zhu, Xiaoyuan and Liu, Kai and Chen, Mian and others},
  title={Wireless Network Digital Twin for 6G: Generative AI as A Key Enabler},
  howpublished={arXiv preprint arXiv:2311.17451},
  year={2023}
}

@misc{grieves2003,
  author={Grieves, Michael},
  title={Digital twin: Manufacturing excellence through virtual copy of physical product},
  year={2003},
  note={Specific publication details vary, but this is the seminal work often cited for the DT concept.}
}

\end{document}